\documentclass[]{rQUF2e}
\usepackage{multirow,setspace,amssymb,amsmath,graphicx,color,rotating,subfigure,url,lineno,natbib}
\usepackage{textcomp}%\usepackage{hyperref}
\usepackage{array}
\newcommand{\PreserveBackslash}[1]{\let\temp=\\#1\let\\=\temp}
\newcolumntype{C}[1]{>{\PreserveBackslash\centering}p{#1}}
\newcolumntype{R}[1]{>{\PreserveBackslash\raggedleft}p{#1}}
\newcolumntype{L}[1]{>{\PreserveBackslash\raggedright}p{#1}}

\bibpunct{(}{)}{,}{a}{}{,}
\begin{document}
\doi{10.1080/1469768YYxxxxxxxx}
 \issn{1469-7696} \issnp{1469-7688} \jvol{00} \jnum{00} \jyear{2008} \jmonth{July}

\markboth{Fei Ren and Wei-Xing Zhou}{\LaTeXe\ guide for authors}

\title{Analysis of trade packages in Chinese stock market}

\author{Fei Ren$\dag$$\ddag$$\S$
\vspace{12pt} and Wei-Xing Zhou$^{\ast}$${\dag}$$\ddag$$\S$\thanks{$^\ast$Corresponding author. Email: wxzhou@ecust.edu.cn}
\\\vspace{12pt}  \normalfont{$\dag$School of Business, East China University of Science and Technology, Shanghai 200237, China\\$\ddag$School of Science, East China University of Science and Technology, Shanghai 200237, China\\$\S$Research Center for Econophysics, East China University of Science and Technology, Shanghai 200237, China}\\\vspace{12pt} }

\maketitle
\begin{abstract}
This paper conducts an empirically study on the trade package composed of a sequence of consecutive purchases or sales of 23 stocks in Chinese stock market. We investigate the probability distributions of the execution time, the number of trades and the total trading volume of trade packages, and analyze the possible scaling relations between them. Quantitative differences are observed between the institutional and individual investors. The trading profile of trade packages is investigated to reveal the preference of large trades on trading volumes and transaction time of the day, and the different profiles of two types of investors imply that institutions may be more informed than individuals. We further analyze the price impacts of both the entire trade packages and the individual transactions inside trade packages. We find the price impact of trade packages is nonnegligible over the period of the execution time and it may have a power-law relation with the total trading volume. The price impact of the transactions inside trade packages displays a U-shaped profile with respect to the time $t$ of the day, and also shows a power-law dependence on their trading volumes. The trading volumes of the transactions inside trade packages made by institutions have a stronger impact on current returns, but the following price reversals persist over a relatively shorter horizon in comparison with those by individuals.
\end{abstract}
\vspace{12pt}

\section{Introduction}
\label{sec:Intro}

The study of price impact of trading on stock exchanges is known as one of the central topics in financial economics. It has been verified by many empirical studies that large trades generally have a strong impact on stock price \citep{Ying-1966-Em,Karpoff-1987-JFQA,Wood-McInish-Ord-1985-JF,Gallant-Rossi-Tauchen-1992-RFS,Chan-Fong-2000-JFE,Lillo-Farmer-Mantegna-2003-Nature,Lim-Coggins-2005-QF,Naes-Skjeltorp-2006-JFinM,Zhou-2011-QF}. The single execution of a large order will lead to a large impact on stock price and increase the investor's cost. Therefore, large orders are usually split into small pieces, and executed at an extended period of time to minimize their price impact. These sequences of trades are called trade packages \citep{Chan-Lakonishok-1995-JF,Gallagher-Looi-2006-AF,Giambona-Golec-2010-JEF}, hidden orders  \citep{Vaglica-Lillo-Moro-Mantegna-2008-PRE,Moro-Vicente-Moyano-Gerig-Farmer-Vaglica-Lillo-2009-PRE,Vaglica-Lillo-Mantegna-2010-NJP}, or metaorders \citep{Farmer-Gerig-Lillo-Waelbroeck-2011-XXX}. Growing evidence shows that large trades play a major role in trading in stock markets, which represent a large fraction of market's total trading volume \citep{Keim-Madhavan-1996-RFS,Jain-2003-JBF,Prino-Jarnecic-Lepone-2007-ABACUS,Gregoriou-2008-JES,Vaglica-Lillo-Mantegna-2010-NJP}.

A large amount of research has been conducted on the large trades of institutional investors, most of which consider the individual trade as the basis unit of analysis in the study of its price impact \citep{Keim-Madhavan-1996-RFS,Prino-Jarnecic-Lepone-2007-ABACUS,Gregoriou-2008-JES,Kraus-Stoll-1972-JF,Aitken-Frino-1996-PBFJ,Gemmill-1996-JF,Saar-2001-RFS,Chiyachantana-Jain-Jiang-Wood-2004-JF}. However, an institutional order is often broken up into a sequence of consecutive trades, and its total volume occupies a large fraction of the stock's trading volume. Therefore, it may be better to treat the sequence of trades as the basis unit of analysis in the study of institutional trades. The earliest study of trade packages may be traced back to the study by \citet{Chan-Lakonishok-1995-JF}. They analyze the price impact of the entire sequence of trades in Now York and American Stock Exchanges, and find that the price impact is related to package size and trade complexity. \citet{Gallagher-Looi-2006-AF} study the abnormal returns for trade packages of the Australian equity managers to estimate their trade performance. \citet{Giambona-Golec-2010-JEF} study the wrong trades inside trade packages by large intuitional insider. Recently, \citet{Vaglica-Lillo-Moro-Mantegna-2008-PRE,Moro-Vicente-Moyano-Gerig-Farmer-Vaglica-Lillo-2009-PRE}, and \citet{Vaglica-Lillo-Mantegna-2010-NJP} try to study the market impact of trade packages identified by the continuous increase or decrease of a firm's stock inventory. \citet{Farmer-Gerig-Lillo-Waelbroeck-2011-XXX} propose a theoretical model to study the permanent impact of large trading orders. The application of such research lies in the study of optimal execution of large orders to minimize the trading costs of institutional investors \citep{Almgren-2003-AMF,Obizhaeva-Wang-2008-JFinM,Alfonsi-Fruth-Schied-2010-QF}.

Many articles have addressed the question of how to measure the price impact of trade packages. \citet{Chan-Lakonishok-1995-JF} study the price impact during trade packages using the average, open and close price on the first and last day of the trade package. Following their works, \citet{Gallagher-Looi-2006-AF} measure the impact over an extended period of time before the start and after the end of the package. Instead, \citet{Vaglica-Lillo-Moro-Mantegna-2008-PRE,Moro-Vicente-Moyano-Gerig-Farmer-Vaglica-Lillo-2009-PRE}, and \citet{Vaglica-Lillo-Mantegna-2010-NJP} study the price impact of trade packages using the transaction data, measured as the difference between the prices of the fist and last transaction of the trade package.

In this paper, we are not only interested in the market impact of the entire trade package, but also concerned about the impact of the individual transactions inside packages. For the study of price impact of isolated large trades, an early study measures the price impact using the close price a few days before and after the trades \citep{Kraus-Stoll-1972-JF}. \citet{Keim-Madhavan-1996-RFS} use the close price on the trading day prior to and after block trades. In stead of using the close price, recent studies choose the transaction price to estimate the price impact of individual trades, e.g., the average price of several transactions prior to and after block trades \citep{Prino-Jarnecic-Lepone-2007-ABACUS,Gemmill-1996-JF}, or the transaction price immediately prior to and after block trades \citep{Gregoriou-2008-JES}.

To the best of our knowledge, the empirical study on trade packages in Chinese stock market has not been conducted. This may due to the difficulty of collection of proprietary data. We have the ultrahigh-frequency data of 23 liquid stocks on the Shenzhen Stock Exchange in year 2003. The data contain the information about all the transactions of each investor, including the trading price, the trading volume, the transaction time, etc. We use the trading records of the 23 stocks as our sample data, and analyze the price impact of trade packages and the individual transactions inside using the transaction price.

Unlike the data provided by the Plexus Group \citep{Keim-Madhavan-1995-RFS,Chiyachantana-Jain-Jiang-Wood-2004-JF}, we are not directly informed of the large institutional trades from our database. Proxy methods are used to detect the trade packages, for instance, a technical detection algorithm based on the inventory time evolution of firms   \citep{Vaglica-Lillo-Moro-Mantegna-2008-PRE,Moro-Vicente-Moyano-Gerig-Farmer-Vaglica-Lillo-2009-PRE,Vaglica-Lillo-Mantegna-2010-NJP}. In the present work, we follow the ideal introduced in \citet{Chan-Lakonishok-1995-JF,Gallagher-Looi-2006-AF} and \citet{Giambona-Golec-2010-JEF}, and define the trade package as a sequence of mostly buy or sell trades with less than a $n$-day break between chronologically adjacent trades. Consistent with the results revealed in previous studies, we observe similar results for various length of break $n=1,5,10$ days, other than some quantitative differences.

So far, previous studies of trade packages primarily focus on the large trades made by institutional investors. Our database includes the trading records for both institutional and individual investors, identified by a particular code denoting the investor type. Large trades made by individuals, e.g., private equity investments, can also be detected using the same detection rules. Therefore, we can also study the price impact of large individual trading. In addition, our present study further confirms the nonlinear power-law relation between the price returns and trading volumes, as revealed in many empirical studies \citep{Hasbrouck-1991-JF,Plerou-Gopikrishnan-Gabaix-Stanley-2002-PRE,Chordia-Subrahmanyam-2004-JFE,Zhou-2011-QF}. Another purpose is to study the price impact of individual transactions inside trade packages. We find that the individual transactions have a strong positive impact on the current price and a temporary negative impact on the following prices, though the cumulative impact of the entire trade package is nonnegligible over the whole period of execution time.

The remainder paper is organized as follows. In Section 2, we introduce our database, the investigated variables of trade packages and their summary statistics. Sections 3 studies the statistic properties of trade packages, including the probability distributions of the investigated variables and their scaling relations. In Section 4, we attempt to study the trading profile of trade packages by analyzing the mean trading volume, transaction probability and the total trading volume with respect to the time of a day. The price impacts of the entire trade packages and the individual transactions inside trade packages are carefully studied in Section 6. Section 7 summarizes our findings.
\vspace{6pt}

\section{Data and investigated variables}
\label{sec:DataVariable}

\subsection{Data sample}
\label{sec:Data}

The data used in our study comprise 23 liquid stocks traded on the Shenzhen Stock Exchange (SZSE), one of the two stock exchanges in mainland China. The SZSE was established on December 1, 1990 and started its operations on July 3, 1991. The SZSE has two separate markets including A-shares and B-shares. A-shares are common stocks issued by mainland Chinese companies, subscribed and traded in Chinese currency {\it Renminbi} (RMB), purchased and sold by Chinese nationals and approved foreign investors. The A-share market was launched in 1990 and opened only to domestic investors before 2003. B-shares are also issued by mainland Chinese companies, but traded in foreign currencies. B-shares carry a face value denominated in RMB. The B-share Market was launched in 1992 and was restricted to foreign investors before February 19, 2001. It has been opened to Chinese investors since then.

We mainly study the data of 23 stocks in A-share market of the SZSE in year 2003. Up to the end of 2003, there have been 491 A-share stocks listed on the SZSE. The total market capitalization of A-share market was 1.2 billion RMB and the float market capitalization was 0.45 billion RMB. Our sample stocks were part of the 40 constituent stocks composing the Shenzhen Stock Exchange Component Index in 2003. These 23 stocks are representative in a variety of industry sectors, as shown in Table~\ref{TB:stock:Summary}. The total transaction amount $A_{tot}$, the float capitalization $C_{flo}$ and the total market capitalization $C_{tot}$, in unit of million RMB, of 23 stocks in year 2003 are also listed in Table~\ref{TB:stock:Summary}. For instance, the total transaction amount $A_{tot}$ of stock 000001 in 2003 is 23847.6 million RMB, and the float capitalization $C_{flo}$ and total market capitalization $C_{tot}$ of the same stock are 12 and 16.6 million RMB respectively. The $A_{tot}$ is three orders of magnitude larger than $C_{flo}$ and $C_{tot}$. Similar phenomenon are observed in all the $23$ stocks, which implies that the average transaction amount per day of a certain stock almost approximates to its total market capitalization. This indicates that the market was relatively active in 2003, though it was in the middle of a five-year bear market \citep{Zhou-Sornette-2004a-PA}.

The SZSE generally opens from Monday to Friday, but closes for public holidays and other dates as announced by the China Securities Regulatory Commission. We mainly consider the continuous double auction operates from 9:30 to 11:30 and 13:00 to 15:00. Based upon the orders of submission and cancelation, the transaction is automatically executed according to price-time priority matching rule. The trading records of 23 stocks are consequently extracting from the original database of order flows. It is worthwhile to point out that the account number for each investor is provided under the condition that the name of the investor involved is removed from the data. Therefore, we can obtain the sequence of transactions for each investor, which contains the information about the trade price, the trading volume, and the transaction time. We treat those transactions executed by a certain investor at the same time but traded with different investors as one trade for this investor. Moreover, in our database each market member is endowed with a code identifying the investor type, i.e., institution and individual. That makes it possible to compare the trading dynamics between those two types of investors.

Table~\ref{TB:stock:Summary} presents summary statistics of the trading records of 23 stocks, including the number of investors $N_{inv}$, the total number of trades $N_{tra}$, the mean, median and standard deviation of the number of trades per investor. The number of investors $N_{inv}$ varies in the range from 34814 (stock 000720) to 533752 (stock 000001), and the total number of trades $N_{tra}$ varies in the range from 194644 (stock 000541) to 2925841 (stock 000001). We integrate the trading records of the 23 stocks, and the sample size is 18597649 trades in total. For most of the stocks analyzed in our study, the mean number of trades for each investor is around 5, and the median is slightly smaller, about 2 or 3. The standard deviation of the number of trades per investor fluctuates in a wide range between 9.3 and 35.4, and the large values of the standard deviation indicate big differences between the trades number of different investors.

\begin{table}
\begin{center}
\begin{minipage}{140mm}
  \tbl{Summary statistics of the trading records of 23 stocks traded on Shenzhen Stock Exchange. The basic information about the 23 stocks includes the stock code and industry they belonging to. We present the summary statistics for each stock, i.e., the total transaction amount $A_{tot}$, the float capitalization $C_{flo}$, the total market capitalization $C_{tot}$, the number of investors $N_{inv}$, the total number of trades $N_{tra}$, the mean, median and standard deviation of the number of trades per investor.}
{\begin{tabular}{crrrrrcccl}
  \toprule
    \multirow{3}*[2mm] & & & & & & \multicolumn{3}{c}{Number of trades per investor} &\\  %
  \cline{7-9}
     Code & $A_{tot}$ & $C_{flo}$ & $C_{tot}$ & $N_{inv}$ & $N_{tra}$ & Mean & Median & Std. dev. & Industry\\
  \colrule
 $000001$ & $23847.6$ & $12.0$ & $16.6$ & $533752$ & $2925841$ & $5.5$ & $3$ & $20.9$ & Financials\\%
 $000002$ & $13024.6$ & $6.1$  & $7.5$  & $299507$ & $1640238$ & $5.5$ & $3$ & $24.4$ & Real estate\\%
 $000009$ & $6287.8$  & $2.6$  & $4.3$  & $261927$ & $1459247$ & $5.6$ & $3$ & $15.2$ & Conglomerates\\%
 $000012$ & $6009.0$  & $0.8$  & $2.8$  & $110038$ & $859938$  & $7.8$ & $3$ & $27.2$ & Metals $\&$ Nonmetals\\%
 $000016$ & $2981.1$  & $1.5$  & $2.7$  & $110641$ & $563163$  & $5.1$ & $2$ & $19.6$ & Electronics\\%
 $000021$ & $8200.8$  & $2.2$  & $8.0$  & $202049$ & $1258294$ & $6.2$ & $3$ & $21.3$ & Electronics\\%
 $000024$ & $2714.6$  & $1.9$  & $4.2$  & $80894$  & $389260$  & $4.8$ & $2$ & $15.6$ & Real estate\\%
 $000027$ & $8486.3$  & $4.4$  & $10.9$ & $218796$ & $1038232$ & $4.7$ & $2$ & $27.2$ & Utilities\\%
 $000063$ & $10436.5$ & $4.7$  & $12.5$ & $127302$ & $799601$  & $6.3$ & $3$ & $35.2$ & IT\\%
 $000066$ & $4255.2$  & $1.5$  & $3.8$  & $144863$ & $845377$  & $5.8$ & $3$ & $14.2$ & Electronics\\%
 $000088$ & $4196.7$  & $3.0$  & $13.9$ & $53420$  & $270488$  & $5.1$ & $2$ & $22.2$ & Transportation\\%
 $000089$ & $5821.2$  & $2.7$  & $7.4$  & $102153$ & $577781$  & $5.7$ & $3$ & $27.4$ & Transportation\\%
 $000406$ & $5197.4$  & $2.2$  & $3.0$  & $152491$ & $832141$  & $5.5$ & $3$ & $14.9$ & Petrochemicals\\%
 $000429$ & $1787.8$  & $1.4$  & $4.7$  & $77888$  & $352044$  & $4.5$ & $2$ & $\ 9.3$& Transportation\\%
 $000488$ & $3876.8$  & $2.4$  & $5.3$  & $54315$  & $347032$  & $6.4$ & $3$ & $35.4$ & Paper $\&$ Printing\\%
 $000539$ & $4840.5$  & $4.2$  & $21.4$ & $61142$  & $317321$  & $5.2$ & $2$ & $20.6$ & Utilities\\%
 $000541$ & $1629.9$  & $1.8$  & $3.3$  & $42290$  & $194644$  & $4.6$ & $2$ & $21.2$ & Electronics\\%
 $000550$ & $7908.4$  & $1.2$  & $5.4$  & $165980$ & $1036314$ & $6.2$ & $3$ & $24.9$ & Manufacturing\\%
 $000581$ & $3031.0$  & $2.5$  & $4.0$  & $60790$  & $283199$  & $4.7$ & $2$ & $22.6$ & Manufacturing\\%
 $000625$ & $14063.8$ & $2.5$  & $13.3$ & $156825$ & $1150636$ & $7.3$ & $3$ & $28.0$ & Manufacturing\\%
 $000709$ & $4155.7$  & $3.0$  & $10.2$ & $138621$ & $650911$  & $4.7$ & $2$ & $23.0$ & Metals $\&$ Nonmetals\\%
 $000720$ & $3684.2$  & $4.9$  & $8.5$  & $34814$  & $332141$  & $9.5$ & $5$ & $16.2$ & Utilities\\%
 $000778$ & $4536.7$  & $2.5$  & $7.0$  & $91802$  & $473806$  & $5.2$ & $2$ & $21.8$ & Manufacturing\\%
   \botrule
  \end{tabular}}
\label{TB:stock:Summary}
\end{minipage}
\end{center}
\end{table}

\subsection{Identification of trade packages}
\label{sec:Ident}

The trade package is defined as a sequence of mostly buy (sell) trades of a stock, and it is generally ended by a specific break time between chronologically adjacent trades. The selection of the break time does not sensitively affect the results, only brings a quantitative difference. \citet{Giambona-Golec-2010-JEF} use an eight-day break to analyze the trade package made by Gabelli asset management company, and they find that the package groupings are not sensitive to the length of the break. In fact, there is no much difference observed for the break length longer than 5 days, as shown in the following context. In this paper, we present the results of trade packages ended by the break of 1,5,10 days separately.

Denote $v_i$ as the trading volume of a certain transaction $i$ inside a trade package, we consider the trade package consisted mostly of purchases or sales, i.e., $\frac{\sum_{buy} v_i}{\sum v_i}>\theta$ or $\frac{\sum_{sell} v_i}{\sum v_i}>\theta$. The parameter $\theta$ is set to be $0.75$ in the present study. For other values $\theta>0.75$ we observe similar results, consistent with the results found by \citet{Vaglica-Lillo-Moro-Mantegna-2008-PRE}. The investors, who occasionally trade in stock market, may have packages containing very few trades which also satisfy the above condition. To avoid considering those investors who are unlikely to split their trades, we choose relatively active investors and restrict the packages with $N_{m}>5$, where $N_{m}$ is the number of trades done through market orders. Market orders are submitted by investors who are urgent to trade, and are executed immediately after the submission. Those orders waited in the limit order book and executed later are limit orders. The threshold 5 is close to the mean number of trades per investor shown in Table~\ref{TB:stock:Summary}. If a package of trades has $N_{m}$ larger than the mean number of trades per investor, we regard it to be a trade package made by an active investor.

Following the detecting rules: (i) packages are separated by a n-day break, (ii) at least $75\%$ of the trading volumes in the package are buying or selling volumes, (iii) the number of trades executed as market orders in the package should be larger than 5, we find out the trade packages of 23 stocks. We integrate the package samples over all the 23 stocks. As shown in Table~\ref{TB:variable:Summary}, there are 1187, 1066, 923 trade packages made by institutions detected for the break of 1, 5, 10 days. For individuals, the number of trade packages detected is larger, 35187, 37924, 35319 for the break of 1, 5, 10 days. This may because that the number of individual investors is much larger than the number of institutional investors. There are 3215246 individuals and 67054 institutions traded in the 23 stocks in year 2003. This may also be explained by their different ways of trading in stock market. Institutions trade strictly following their trading strategies, and they do not trade as frequently as individuals.

The variables characterizing the trade packages are defined as following: (i) The execution time $T$ of the trade package, measured as the interval between the first and last transaction of the trade package, in unit of one second. (ii) The number of trades $N$ within the trade package. (iii) The total trading volume of the trade package, denoted as $V=\sum_{i=1}^{N} v_i$.

Table~\ref{TB:variable:Summary} gives the means of the execution time, the number of trades and the total trading volumes for both institutions and individuals. The mean of the execution time $\langle T \rangle$ shows an increasing tendency as the increase of the break time separating packages. It is 5973, 57259, 113779 seconds for institutions, and 5553, 58701, 117811 seconds for individuals for the break of 1, 5, 10 days. The mean of the number of trades $\langle N \rangle$ for institutions is 41, 54, 56 for the break of 1, 5, 10 days, larger than 32, 35, 36 those for individuals. Similarly, the mean of the total trading volumes $\langle V \rangle$ for institutions is 177311, 224872, 236013 for the break of 1, 5, 10 days, larger than 121841, 124456, 124004 those for individuals. This implies the institutions have capitalization on average larger than individuals, and need more transactions to accomplish the purchase or sale of large amounts of shares. In general, a trade package consists of a sequence of $30-60$ consecutive transactions, and each transaction on average has $4000$ shares. Therefore, the trade packages analyzed in our study are of extremely large size.

\section{Statistics properties of trade packages}
\label{sec:Stat}

\subsection{Probability distributions of variables characterizing trade packages}
\label{sec:PDF}

\citet{Vaglica-Lillo-Moro-Mantegna-2008-PRE} study three most capitalized stocks traded on the Spanish Stock Exchange (BME), and reveal that the probability distribution functions (PDFs) of the variables $T$, $N$ and $V$ have power-law tails with exponents lying in an interval $[1.2,2.3]$. An extended study of 23 stocks on BME and 74 stocks on the London Stock Exchange (LSE) further confirms this result \citep{Moro-Vicente-Moyano-Gerig-Farmer-Vaglica-Lillo-2009-PRE}. We pool together the data from 23 stocks on the SZSE, and investigate the PDFs of $T$, $N$ and $V$ for trade packages ended by the break of 1,5,10 days. Power-law tails are observed in the PDFs of these variables. For a variable $x$ which obeys a power-law tail distribution, we suppose its PDF follows a formula
\begin{equation}
  P(x)=c x^{-\delta},\ x\geq x_{\min}.
  \label{Eq:PL}
\end{equation}

\begin{figure}
\begin{center}
\begin{minipage}{100mm}
\subfigure[]{
\resizebox*{5cm}{!}{\includegraphics{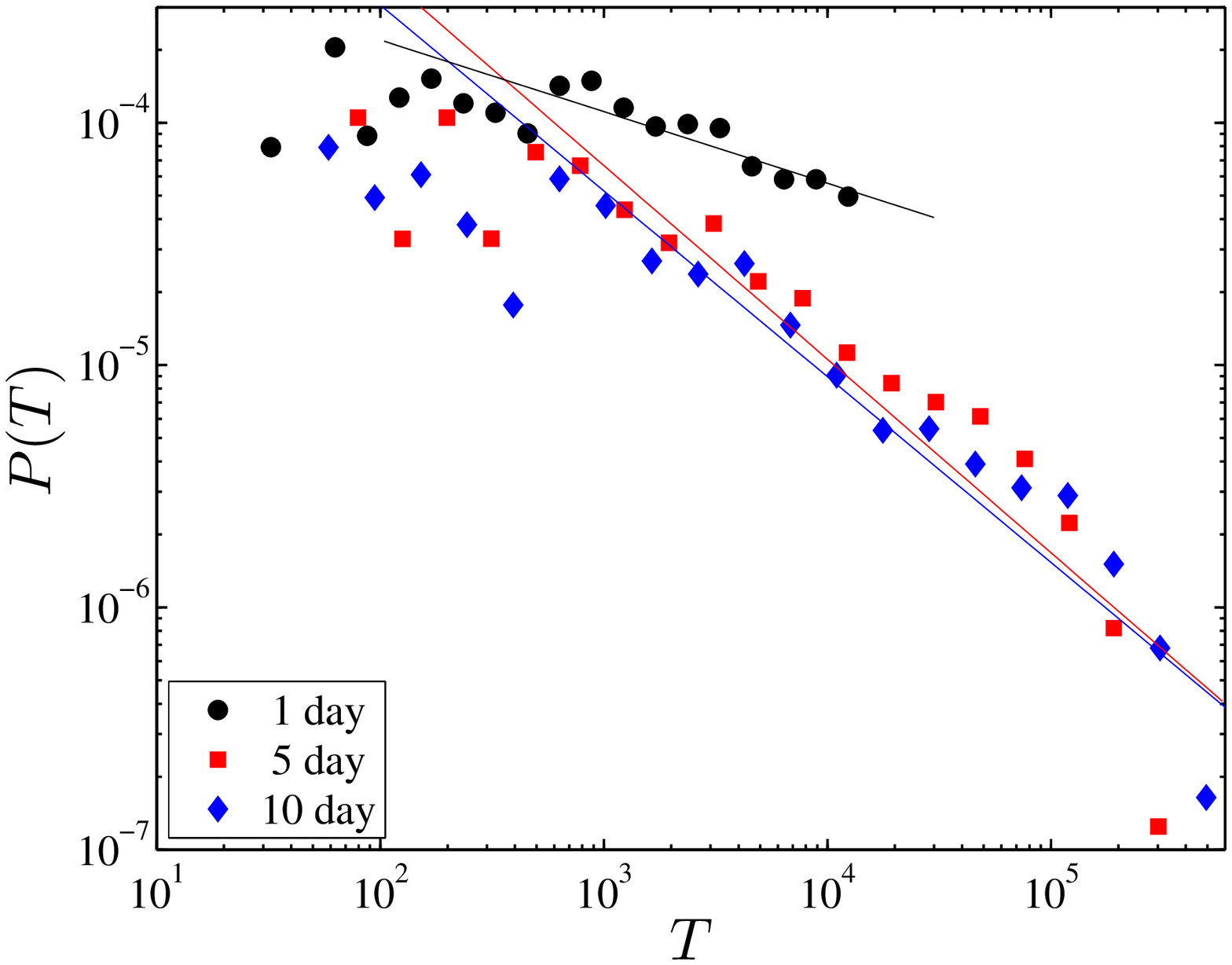}}\label{PDF-TNV_part_a}}%
\subfigure[]{
\resizebox*{5cm}{!}{\includegraphics{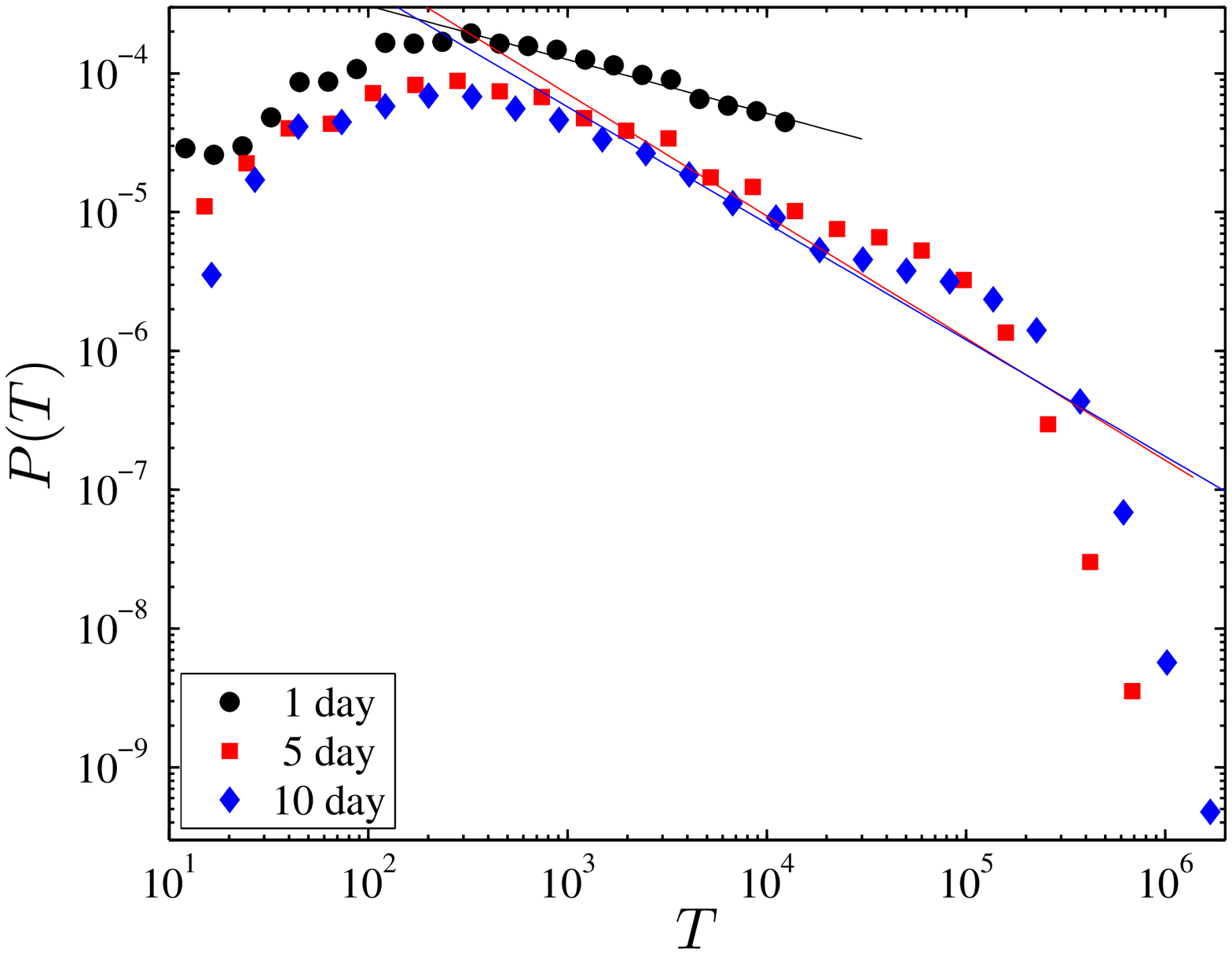}}\label{PDF-TNV_part_b}}%
\\
\subfigure[]{
\resizebox*{5cm}{!}{\includegraphics{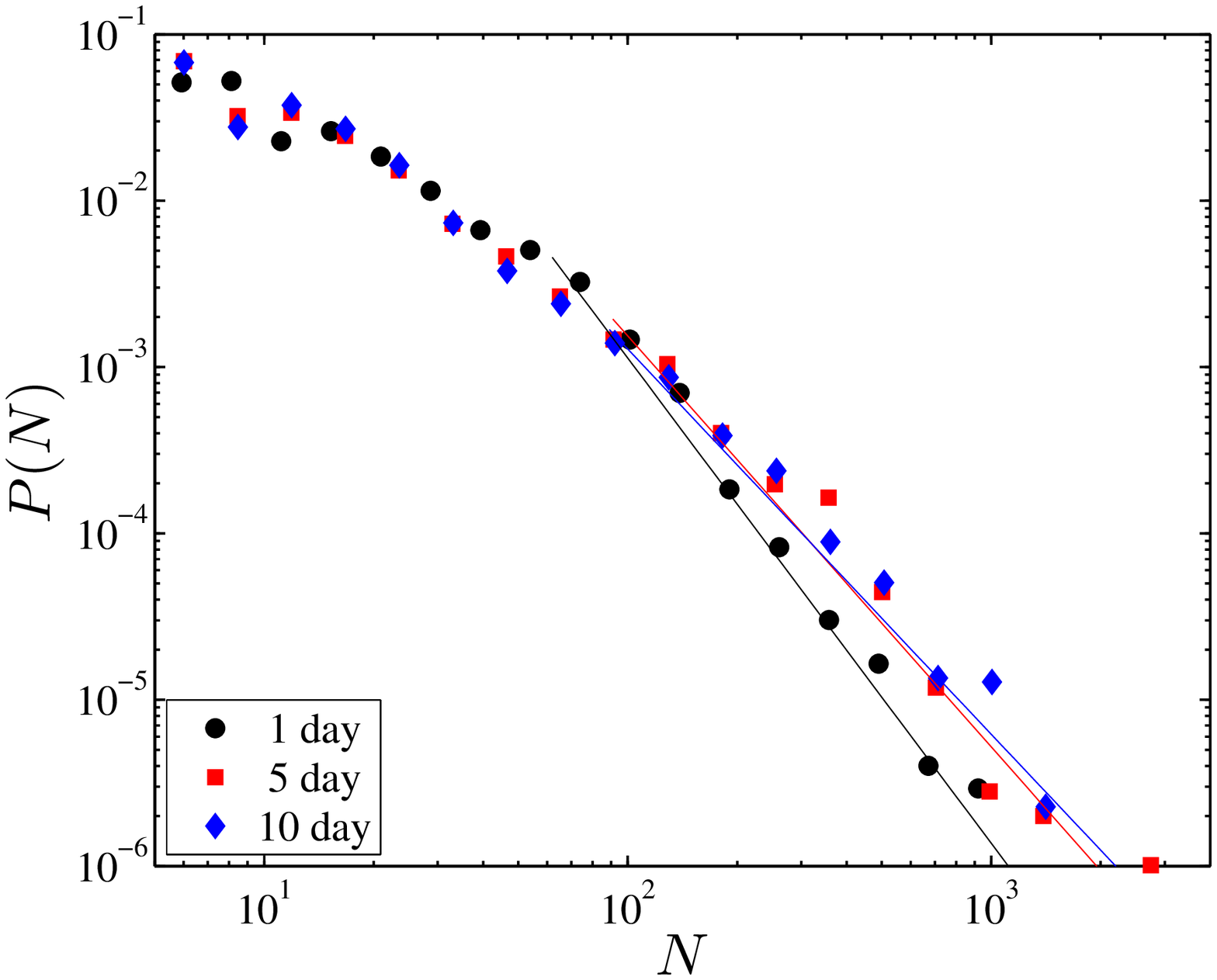}}\label{PDF-TNV_part_c}}%
\subfigure[]{
\resizebox*{5cm}{!}{\includegraphics{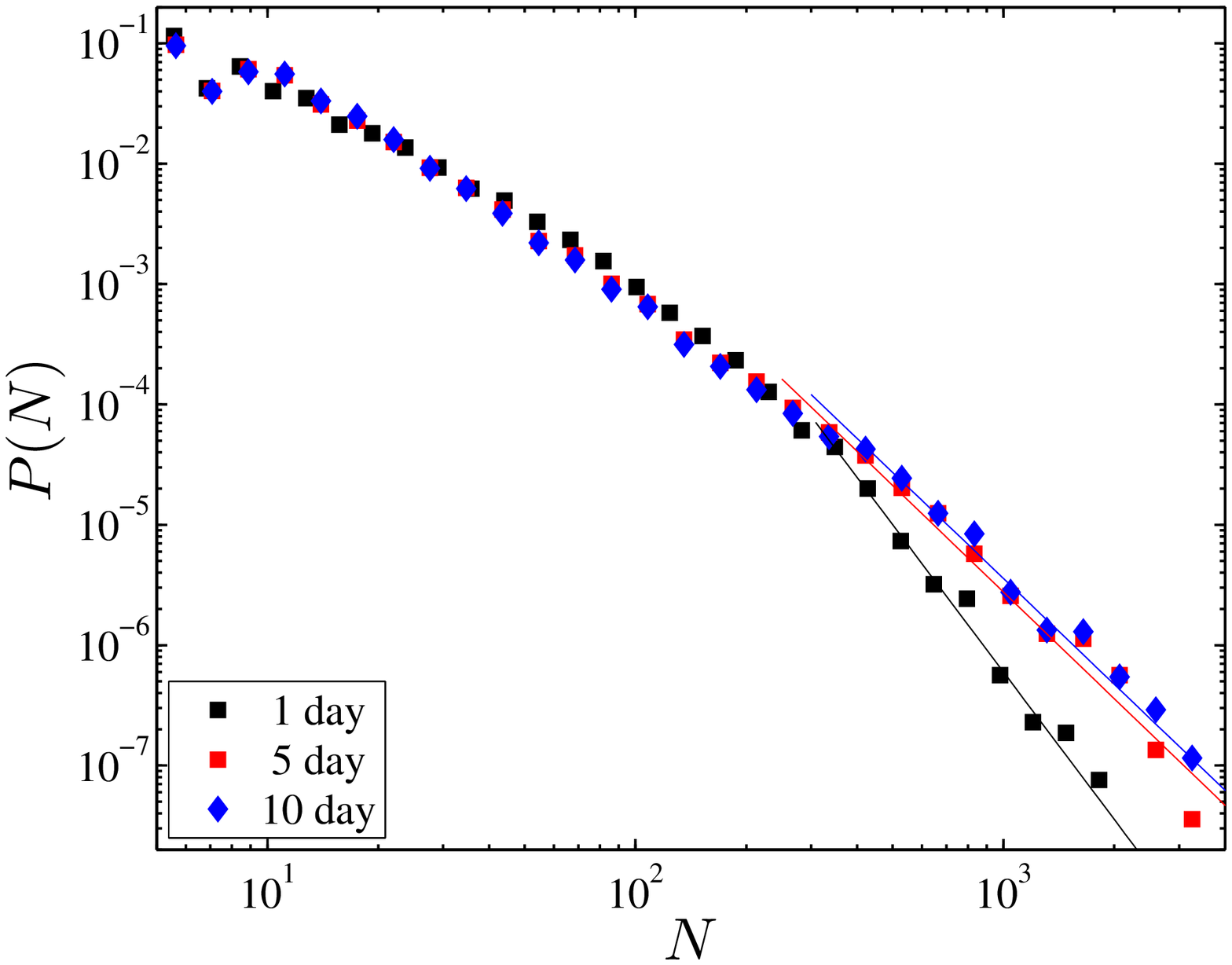}}\label{PDF-TNV_part_d}}%
\\
\subfigure[]{
\resizebox*{5cm}{!}{\includegraphics{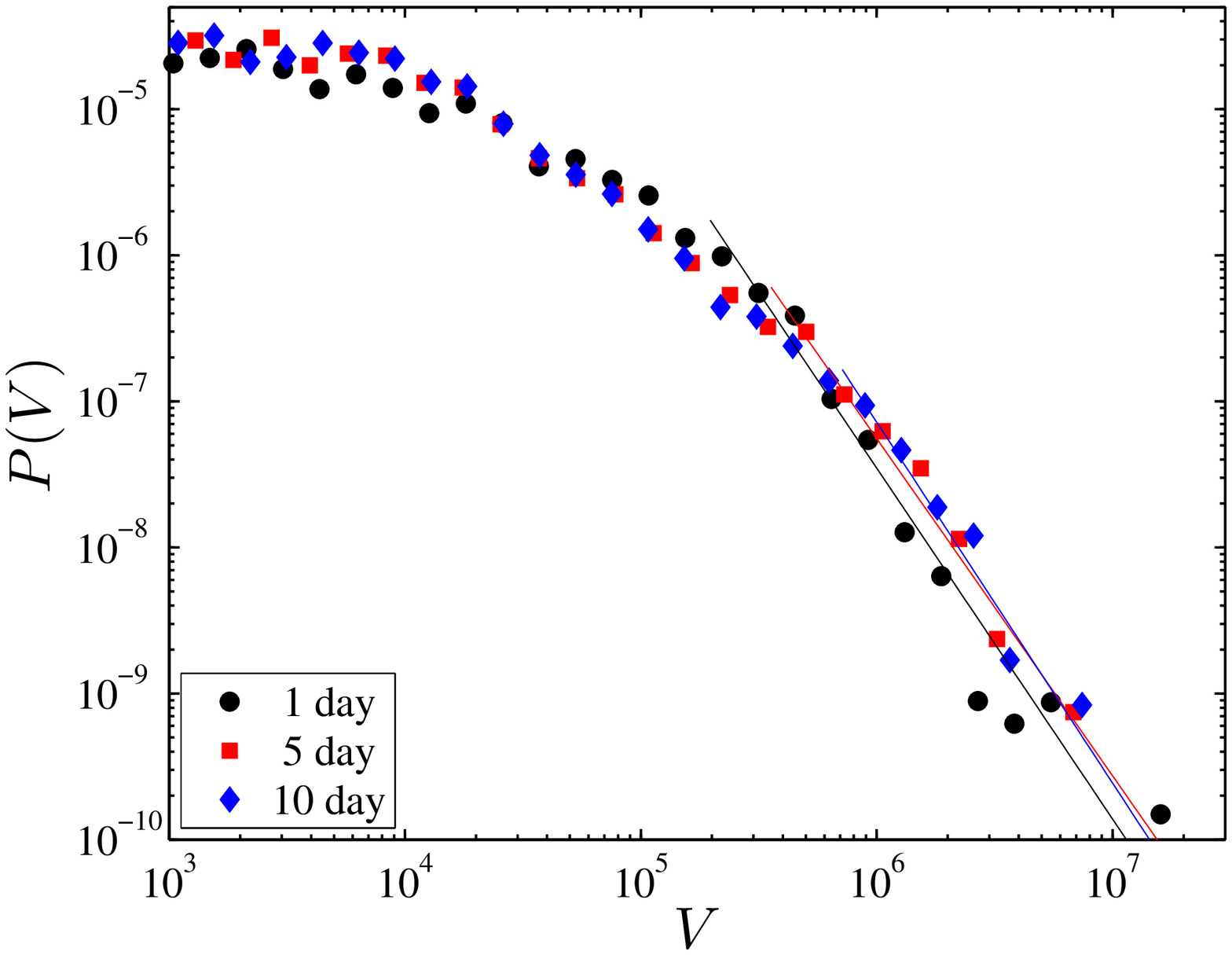}}\label{PDF-TNV_part_e}}%
\subfigure[]{
\resizebox*{5cm}{!}{\includegraphics{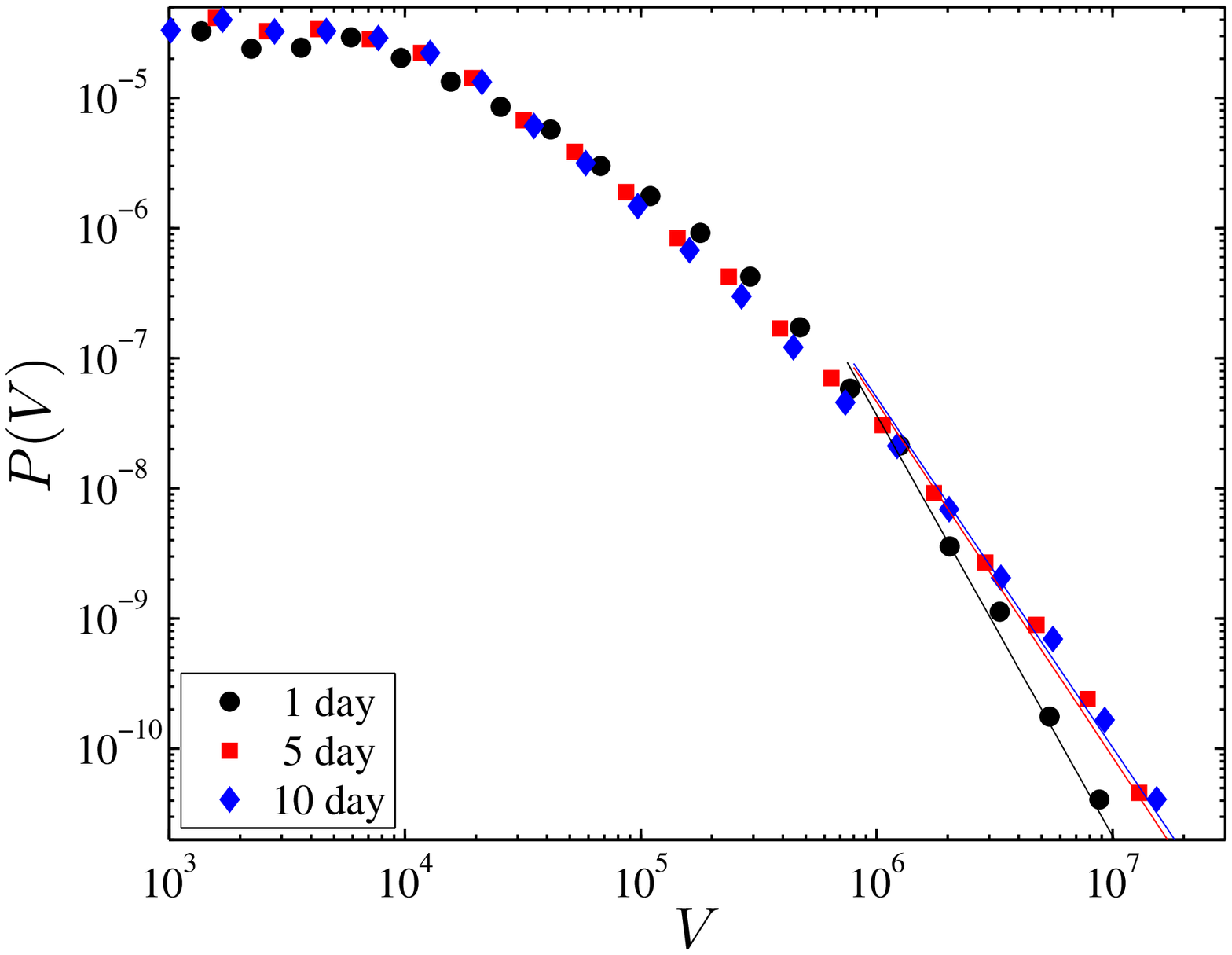}}\label{PDF-TNV_part_f}}%
\label{PDF-TNV}
\end{minipage}
\end{center}
\caption{(Color online) Probability distribution functions (PDFS) of the variables characterizing the trade packages with various break length $n=1,5,10$ days using the aggregated data from the 23 stocks: (a) PDF of the execution time $T$ for institutions, (b) PDF of the execution time $T$ for individuals, (c) PDF of the number of trades $N$ for institutions, (d) PDF of the number of trades $N$ for individuals, (e) PDF of the trading volume $V$ for institutions, and (d) PDF of the trading volume $V$ for individuals. The solid curves are power-law fits with exponents depicted in Table~\ref{TB:variable:Summary}.}
\end{figure}

We first calculate the PDF of the execution time $T$. Note the PDF is different for different types of investors, we present the results for institutions and individuals respectively. In Figures~\ref{PDF-TNV_part_a} and ~\ref{PDF-TNV_part_b}, $P(T)$ of trade packages ended by the break of 1,5,10 days for both institutions and individuals are plotted. The PDFs seem to have power-law tails, and we fit them using a power-law function presented in Equation~(\ref{Eq:PL}). A rough estimation shows the exponent $\delta$ might smaller than one. For a sequence of variable $x=x_1,\cdots,x_n$ restricted to a finite region $[x_{\min},x_{\max}]$, the parameter $c$ is
\begin{equation}
  c=\frac{1-\delta}{x_{\max}^{1-\delta} - x_{\min}^{1-\delta}},
  \label{Eq:PL:C}
\end{equation}
under the normalization condition $\int_{x_{\min}}^{x_{\max}} p(x)dx=1$. To obtain the maximum likelihood estimate of the exponent $\delta$, one needs to solve the equation
\begin{equation}
  \frac{\partial L}{\partial \delta}=\frac{n}{\delta-1}-n\frac{x_{\min} \ln x_{\min} x_{\max}^\delta - x_{\max} \ln x_{\max} x_{\min}^\delta }{x_{\max} x_{\min}^\delta -x_{\min} x_{\max}^\delta}-\sum_{i=1}^{n}x_i=0,
  \label{Eq:PL:L}
\end{equation}

Following the work by \citet{Clauset-Shalizi-Newman-2009-SIAMR}, we use an efficient method of fitting power-law distributions based on the Kolmogorov-Smirnov (KS) statistic to
estimate the parameters in Equation~(\ref{Eq:PL}). The $KS$ statistic is defined as
\begin{equation}
   KS = \max_{x\geq \hat{x}_{\min}} \left(|F-F_{\rm{PL}}|\right),
   \label{Eq:KS}
\end{equation}
where $F$ is the cumulative distribution function (CDF) of the empirical data and $F_{\rm{PL}}$ is the CDF of the power-law fit.
The CDF of the power-law function in Equation~(\ref{Eq:PL}) with parameter $c$ in Equation~(\ref{Eq:PL:C}) is
\begin{equation}
  F_{\rm{PL}}(x)=1-\frac{1-x_{\min}^{\delta-1}x^{1-\delta}}{1-x_{\min}^{\delta-1}x_{\max}^{1-\delta}}.
  \label{Eq:PL:CDF}
\end{equation}
To make the empirical PDF and the best power-law fit as similar as possible, we determine the estimate of $\hat{x}_{\min}$ by minimizing
the $KS$ statistic, then we estimate the exponent $\delta$ by solving Equation~(\ref{Eq:PL:L}). The parameter $c$ is obtained by substituting $\hat{x}_{\min}$ and $\delta$ into Equation~(\ref{Eq:PL:C}). The standard error on $\delta$, according to \citet{Clauset-Shalizi-Newman-2009-SIAMR}, is
\begin{equation}
   \sigma = -\frac{1}{n E \left[\frac{\partial^2 L}{\partial \delta^2} \right]}.
   \label{Eq:Err}
\end{equation}

As shown in Figure~\ref{PDF-TNV_part_a}, $P(T)$ for institutions displays power-law tails covering a scaling range over two orders of magnitude, and the exponent $\delta_T$ varies with different break length. For the break of one day, $\delta_T$ is estimated to be $0.3$, while for the break longer than five days $\delta_T$ is relatively constant and is around $0.8$. In the figure, the power-law fits with parameters listed in Table~\ref{TB:variable:Summary} are also illustrated. For individuals, we observe similar results, and the estimated values of $\delta_T$ are slightly larger than those for the institution, as depicted in Table~\ref{TB:variable:Summary}.

We then calculate the PDF of the number of trades $N$. In Figures~\ref{PDF-TNV_part_c} and ~\ref{PDF-TNV_part_d}, we plot $P(N)$ of trade packages ended by the break of 1,5,10 days for both institutions and individuals. It also has power-law tails, but with exponent $\delta_N$ obviously greater than 1.0. For $\delta>1.0$, the parameter $c$ in Equation~(\ref{Eq:PL}) in the limit of $x_{max}\rightarrow \infty$ approximates
\begin{equation}
  c=\frac{\delta-1}{x_{\min}^{1-\delta}},
  \label{Eq:PL:C1}
\end{equation}
and the exponent is
\begin{equation}
  \hat{\delta}=1+n \left( \sum_{i=1}^{n} \ln \frac{x_i}{x_{\min}} \right)^{-1}.
  \label{Eq:PL:exp}
\end{equation}
This formula is the same as that proposed by \citet{Clauset-Shalizi-Newman-2009-SIAMR}. We use the same power-law fitting method based on KS statistic to estimate the parameters $\hat{x}_{\min}$ and $\delta$. Now the CDF of the power-law fit with $\delta>1.0$ is
\begin{equation}
  F_{\rm{PL}}(x)=1- \left( \frac{x_{\min}}{x} \right) ^{\delta-1}.
  \label{Eq:PL:CDF1}
\end{equation}
The standard error on $\delta$ is obtained according to Equation~(\ref{Eq:Err}).

The power-law exponent $\delta_N$ varies with different break length. For institutions, $\delta_N$ is estimated to be $2.9$ for the break of one day, and about $2.4$ for the break longer than five days. For individuals, $\delta_N$ is $4.0$ for the break of one day, and about $2.9$ for the break longer than five days. The power-law fits are also illustrated in the figure, and the estimated values of $\delta_N$ are depicted in Table~\ref{TB:variable:Summary}.

We also calculate the PDF of the total trading volume $V$, and similar power-law tails are observed. In Figures~\ref{PDF-TNV_part_e} and ~\ref{PDF-TNV_part_f}, $P(V)$ of trade packages ended by the break of 1,5,10 days for both institutions and individuals are plotted. For institutions, $\delta_V$ for the break of one day is $2.4$, displaying a value very close to that for the break longer than five days, while for individuals $\delta_V$ for the break of one day is $3.2$ larger than $2.7$ that for the break longer than five days. The power-law fits of $P(V)$ with parameters listed in Table~\ref{TB:variable:Summary} are also illustrated in the figure.

\begin{table}
\begin{center}
\begin{minipage}[c]{0.8\linewidth}
  \tbl{Summary properties of trade packages of the aggregated data from the 23 stocks. The results are presented for trade packages made by institutions and individuals with various break length $n=1,5,10$ days. $N_p$ is the number of packages, $\langle T \rangle$, $\langle N \rangle$ and $\langle V \rangle$ are means of the execution time, the number of trades and the total trading volumes. The power-law exponents $\delta_T$, $\delta_N$ and $\delta_V$ are obtained by fitting Equation~(\ref{Eq:PL}), estimated by a maximum likelihood method based on $KS$ statistic. The exponents $g_1$, $g_2$ and $g_3$ describe the scaling relations in Equation~(\ref{Eq:Sca:Rel}).}
{\begin{tabular}{cccccccc}
  \toprule
    \multirow{3}*[2mm]  & \multicolumn{3}{c}{Institution} & & \multicolumn{3}{c}{Individual}\\  %
  \cline{2-4} \cline{6-8}
     & $1$ day & $5$ days & $10$ days & & $1$ day & $5$ days & $10$ days \\
  \colrule
   $N_p$ & $1187$ & $1066$ &  $923$ & & $35187$ & $37924$ & $35319$\\
   $\langle T \rangle$ & $5973$ & $57259$ &  $113779$ & & $5553$ & $58701$ & $117811$\\
   $\langle N \rangle$ &   $41$ & $54$ &  $56$ & & $32$ & $35$ & $36$\\
   $\langle V \rangle$ & $177311$ & $224872$ &  $236013$ & & $121841$ & $124456$ & $124004$\\
   $\delta_T$ & $0.30\pm0.03$ & $0.80\pm0.01$ &  $0.77\pm0.01$ & & $0.39\pm0.005$ & $0.88\pm0.002$ & $0.84\pm0.002$\\
   $\delta_N$ & $2.92\pm0.13$ & $2.47\pm0.12$ &  $2.31\pm0.11$ & & $4.06\pm0.19$  & $2.94\pm0.09$  & $2.92\pm0.09$ \\
   $\delta_V$ & $2.40\pm0.08$ & $2.31\pm0.10$ &  $2.47\pm0.17$ & & $3.23\pm0.08$  & $2.73\pm0.09$  & $2.69\pm0.09$ \\
   $g_1$ & $0.18\pm0.05$ & $0.49\pm0.08$ & $0.49\pm0.16$ & & $0.13\pm0.03$ & $0.38\pm0.05$ & $0.34\pm0.06$ \\
   $g_2$ & $0.74\pm0.08$ & $0.77\pm0.05$ & $0.77\pm0.07$ & & $0.72\pm0.07$ & $0.79\pm0.08$ & $0.80\pm0.08$ \\
   $g_3$ & $0.18\pm0.08$ & $0.51\pm0.17$ & $0.53\pm0.15$ & & $0.16\pm0.03$ & $0.56\pm0.07$ & $0.43\pm0.04$ \\
   \botrule
  \end{tabular}}
\label{TB:variable:Summary}
\end{minipage}
\end{center}
\end{table}

\subsection{Scaling relations between variables characterizing trade packages}
\label{sec:Scal}

Empirical studies have revealed that there exist scaling relations between variables $T$, $N$ and $V$
\begin{equation}
  T \sim V^{g_1}, \quad N \sim V^{g_2}, \quad T \sim N^{g_3}.
  \label{Eq:Sca:Rel}
\end{equation}
It is found $g_1\simeq2$, $g_2\simeq1$, and $g_3\simeq1.7$ for BME and LSE markets \citep{Moro-Vicente-Moyano-Gerig-Farmer-Vaglica-Lillo-2009-PRE}. We also find that $T$, $N$ and $V$ of the trade packages for SZSE are related through scaling relations, but with different values of $g_1$, $g_2$ and $g_3$, primarily due to the small value of the exponent $\delta_T$. Since the exponents $\delta_T$, $\delta_N$ and $\delta_V$ differ for the two types of investors, we calculate the scaling relations for institutions and individuals respectively.

\begin{figure}
\begin{center}
\begin{minipage}{120mm}
\subfigure[]{
\resizebox*{4cm}{!}{\includegraphics{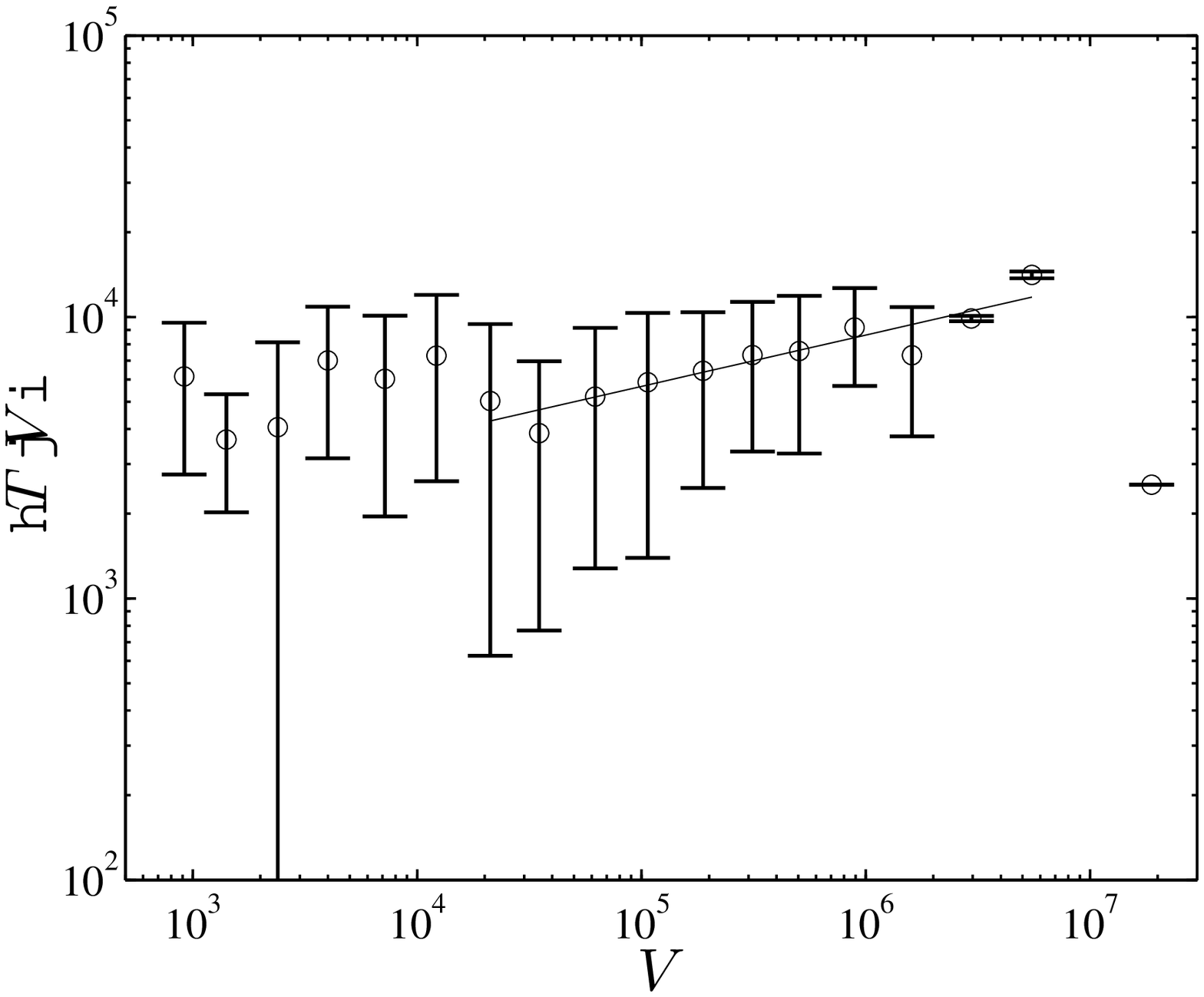}}\label{ScaRel-TNV_part_a}}%
\subfigure[]{
\resizebox*{4cm}{!}{\includegraphics{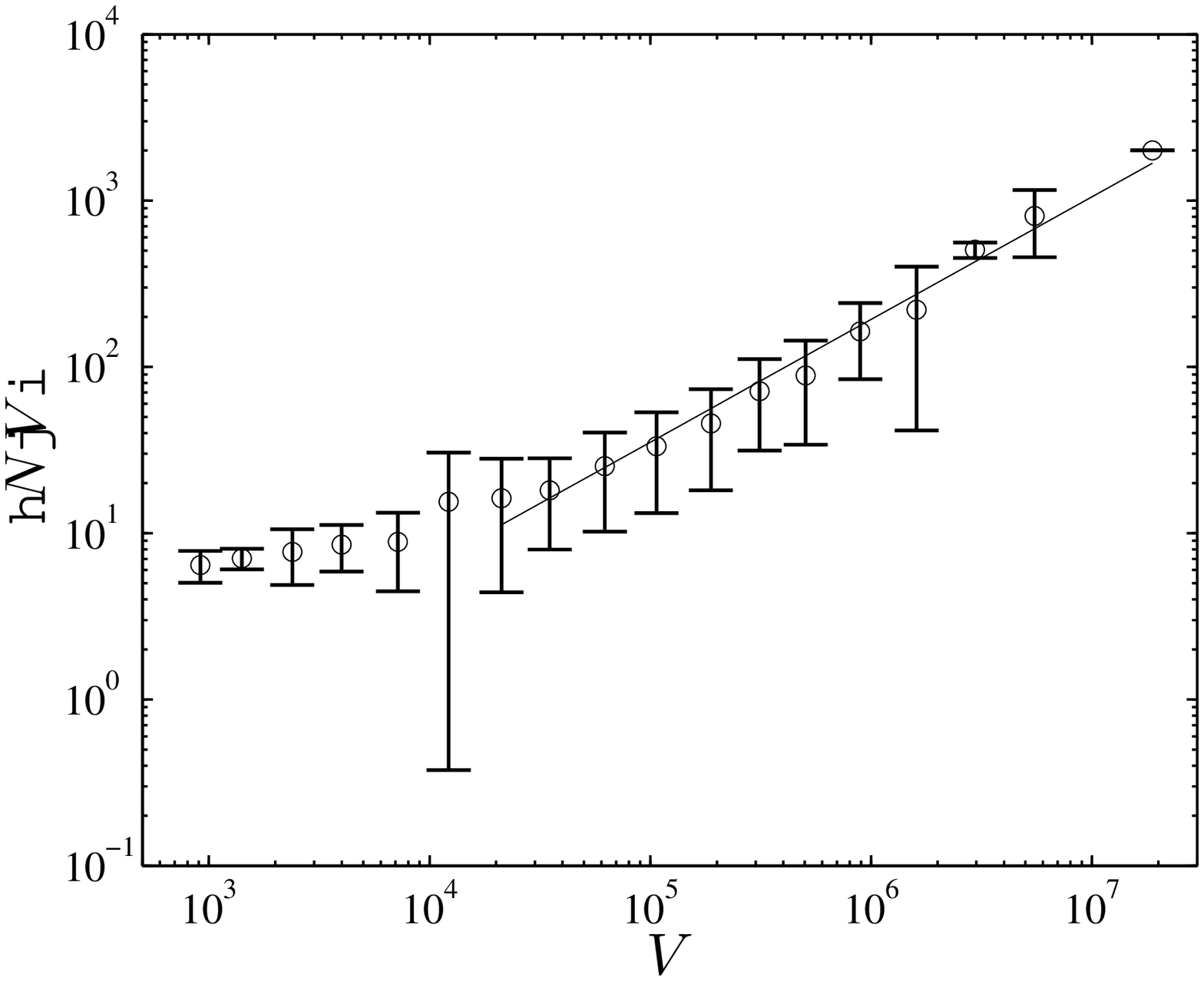}}\label{ScaRel-TNV_part_b}}%
\subfigure[]{
\resizebox*{4cm}{!}{\includegraphics{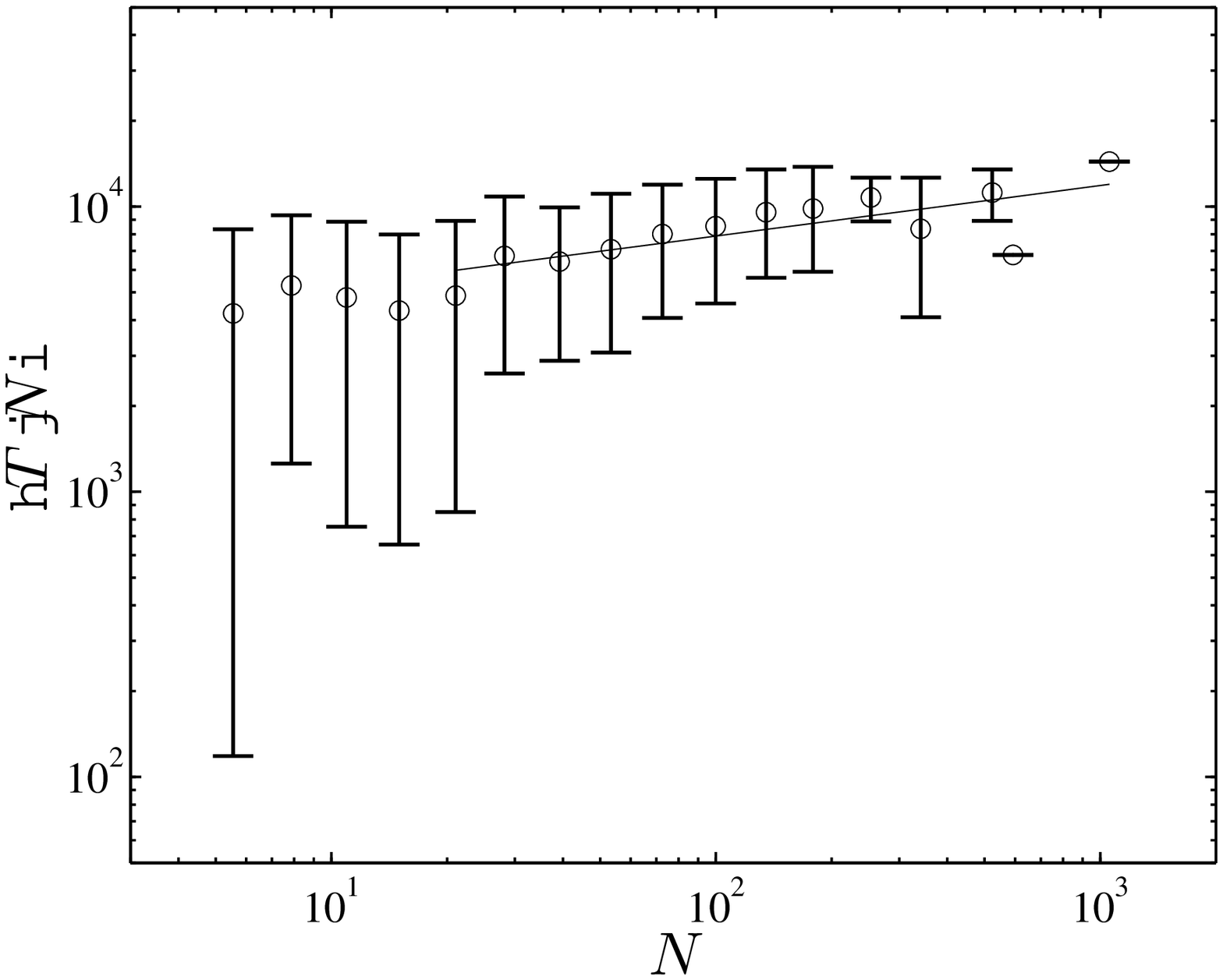}}\label{ScaRel-TNV_part_c}}%
\\
\subfigure[]{
\resizebox*{4cm}{!}{\includegraphics{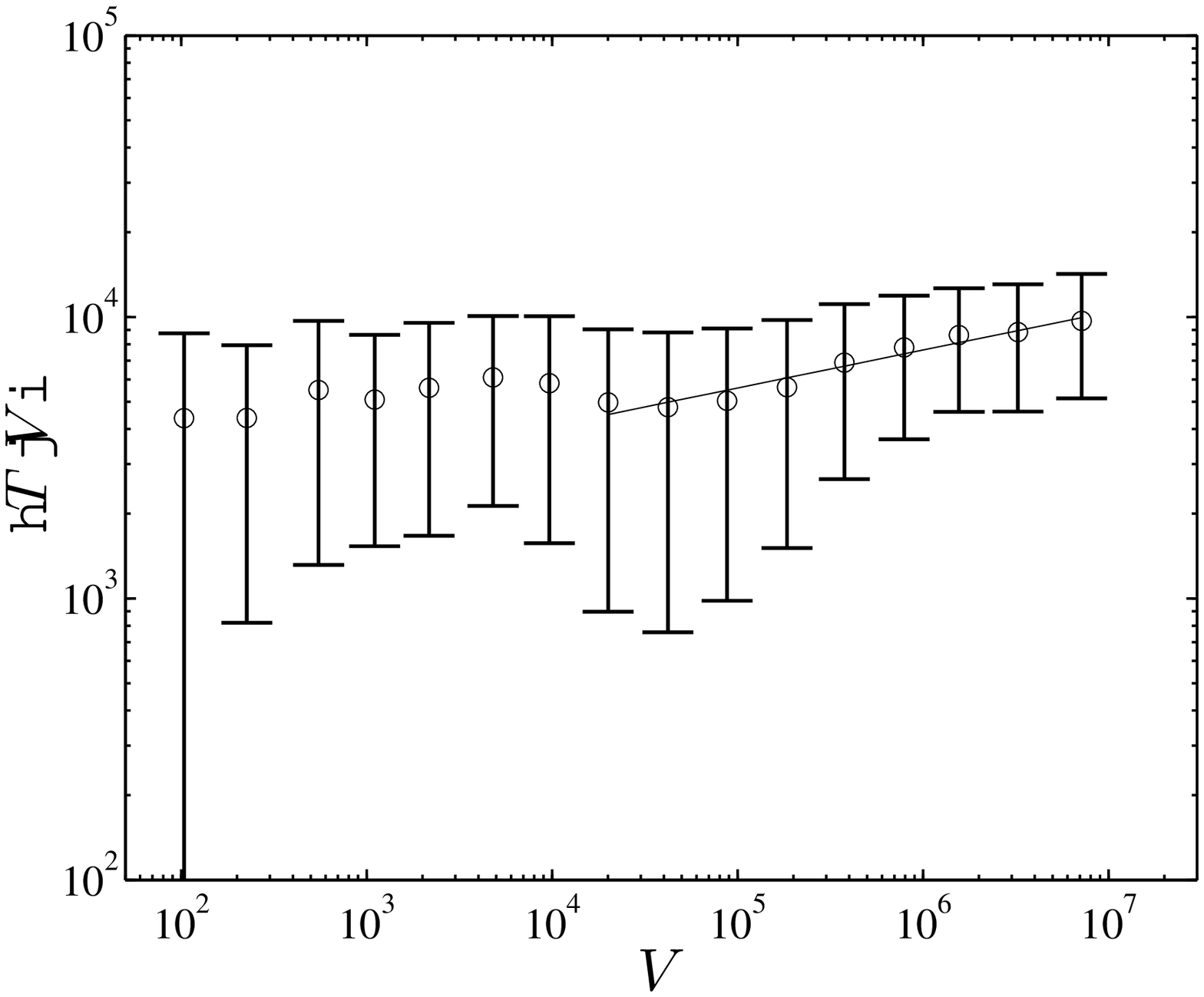}}\label{ScaRel-TNV_part_d}}%
\subfigure[]{
\resizebox*{4cm}{!}{\includegraphics{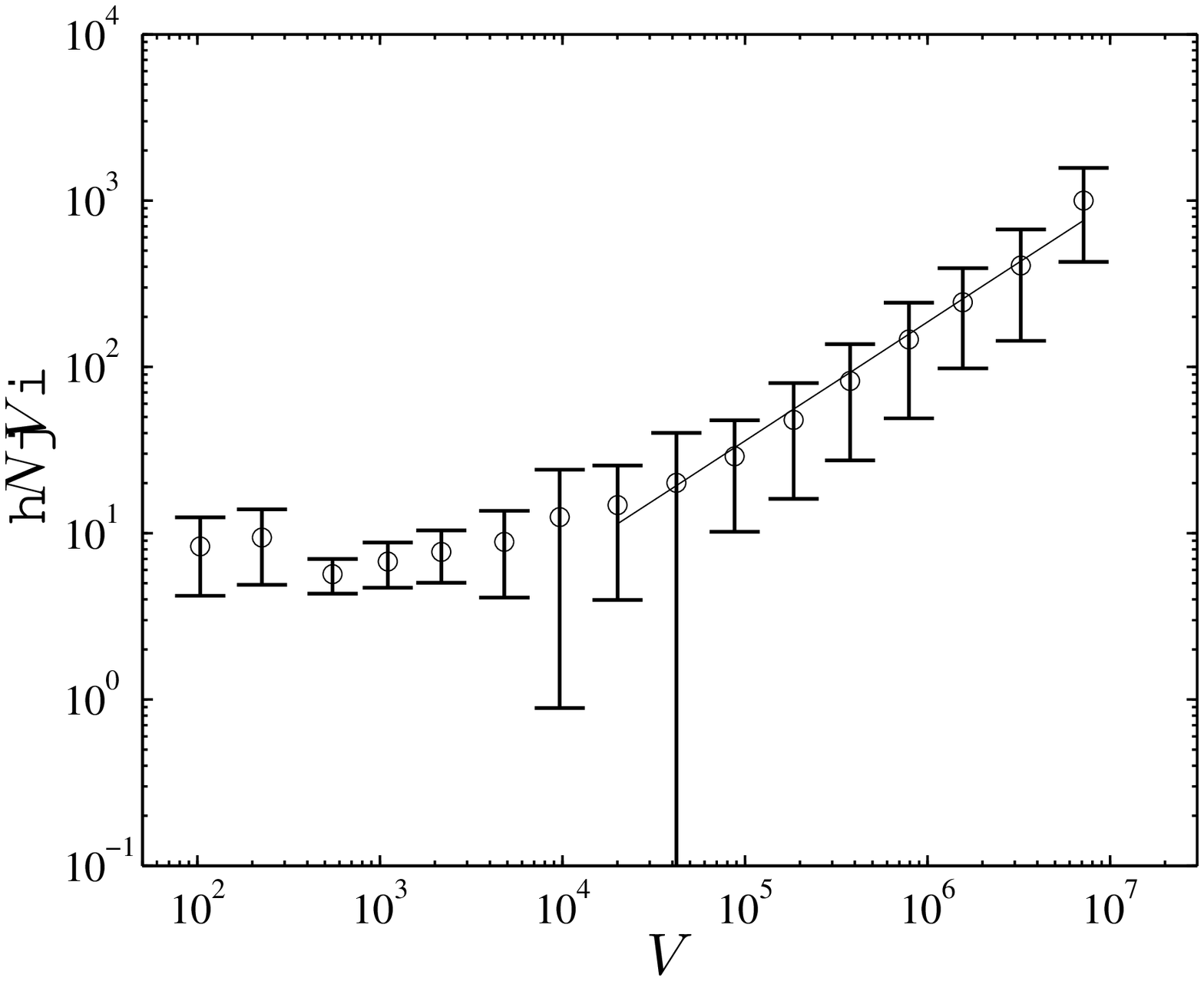}}\label{ScaRel-TNV_part_e}}%
\subfigure[]{
\resizebox*{4cm}{!}{\includegraphics{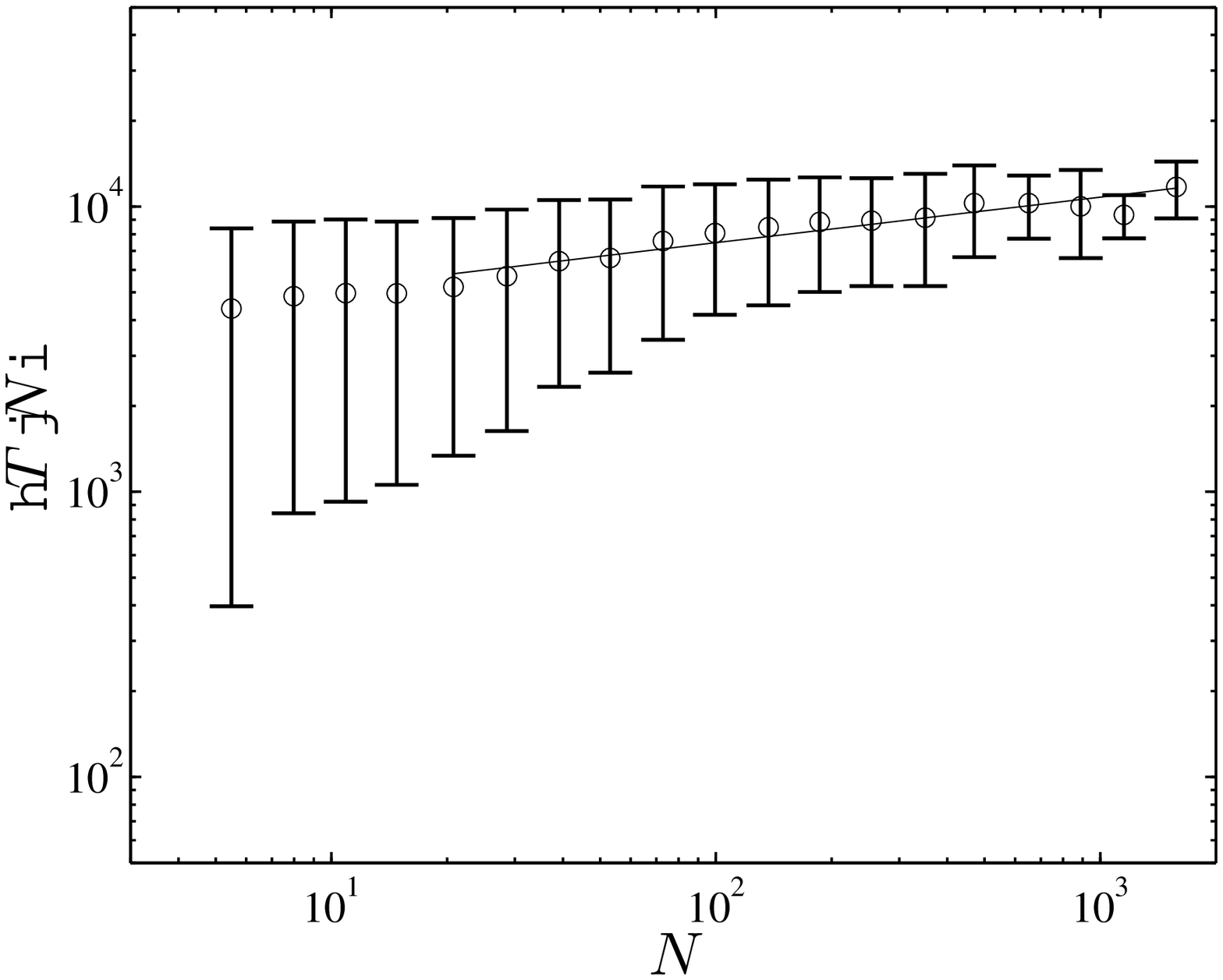}}\label{ScaRel-TNV_part_f}}%
\label{ScaRel-TNV}
\end{minipage}
\end{center}
\caption{(Color online) Scaling relations between the variables characterizing the trade packages with break length $n=1$ day: (a) mean conditional execution time $\langle T|V \rangle$ for institutions, (b) mean conditional number of trades $\langle N|V \rangle$ for institutions, (c) mean conditional execution time $\langle T|N \rangle$ for institutions, (d) mean conditional execution time $\langle T|V \rangle$ for individuals, (e) mean conditional number of trades $\langle N|V \rangle$ for individuals, and (f) mean conditional execution time $\langle T|N \rangle$ for individuals. The solid curves are power-law fits with exponents depicted in Table~\ref{TB:variable:Summary}.}
\end{figure}

We study the scaling relation between $T$ and $V$ by calculating the mean conditional execution time $\langle T|V \rangle$ conditioned on the total trading volume $V$. We arrange the entire $T$ sequences in ascending order, and partition it to 20 bins with equal size. We calculate $\langle T|V \rangle$ conditioned on a bin of $V$ to get better statistics. In Figure~\ref{ScaRel-TNV_part_a}, $\langle T|V \rangle$ for trade packages of institutions ended by the break of one day is plotted. For large scales of $V$, one observes a scaling relation between $T$ and $V$, and $g_1$ is estimated to be $0.18$. This is consistent with the previous result that $P(V)$ obeys power law for large $V$. We also study the scaling relations between $N$ and $V$, $T$ and $N$ using the same method, and obtain $g_2=0.74$ and $g_3=0.18$ as illustrated in Figures~\ref{ScaRel-TNV_part_b} and ~\ref{ScaRel-TNV_part_c}. For trade packages of individuals ended by the break of one day, we obtain $g_1=0.13$, $g_2=0.72$ and $g_3=0.16$, slightly smaller than those for institutions, as shown in Figures~\ref{ScaRel-TNV_part_d}, ~\ref{ScaRel-TNV_part_e} and ~\ref{ScaRel-TNV_part_f}. Similar results are observed for trade packages of both institutions and individuals ended by the break longer than five days. According to the scaling relations in Equation~(\ref{Eq:Sca:Rel}), one can infer the relation that $g_1=g_2 g_3$. The estimated values of $g_1$, $g_2$ and $g_3$ depicted in Table~\ref{TB:variable:Summary} prove the inferred relation between them.

\section{Trading profile}
\label{sec:Profile}

We further consider the question of how the individual transactions inside trade packages are executed as a function of the time $t$ of a day, which is called trading profile. We are mainly concerned about the trade packages finished within one day, i.e., packages ended by the break of one day, and the time $t$ is normalized by the time of a trading day $D=14400$ seconds. Denote $v_i$ as the trading volume of transaction $i$ inside trade packages, we calculate the mean volume of each transaction $\langle v(t) \rangle$ traded at normalized time $t/D$. To make the trading volumes of different stocks comparable, $v_i$ of a particular stock is normalized by its mean value.

\begin{figure}
\begin{center}
\begin{minipage}{100mm}
\subfigure[]{
\resizebox*{5cm}{!}{\includegraphics{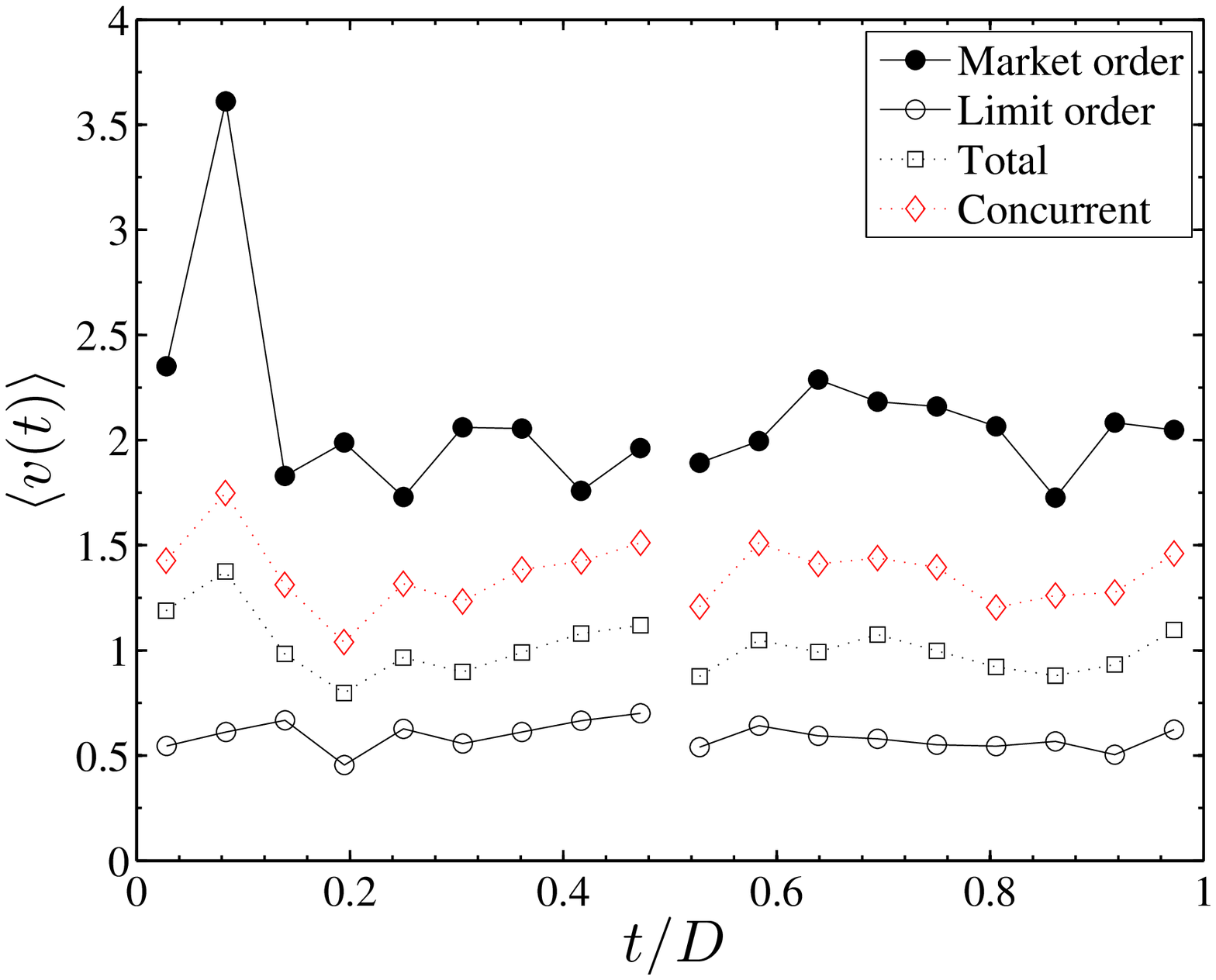}}\label{Profile_part_a}}%
\subfigure[]{
\resizebox*{5cm}{!}{\includegraphics{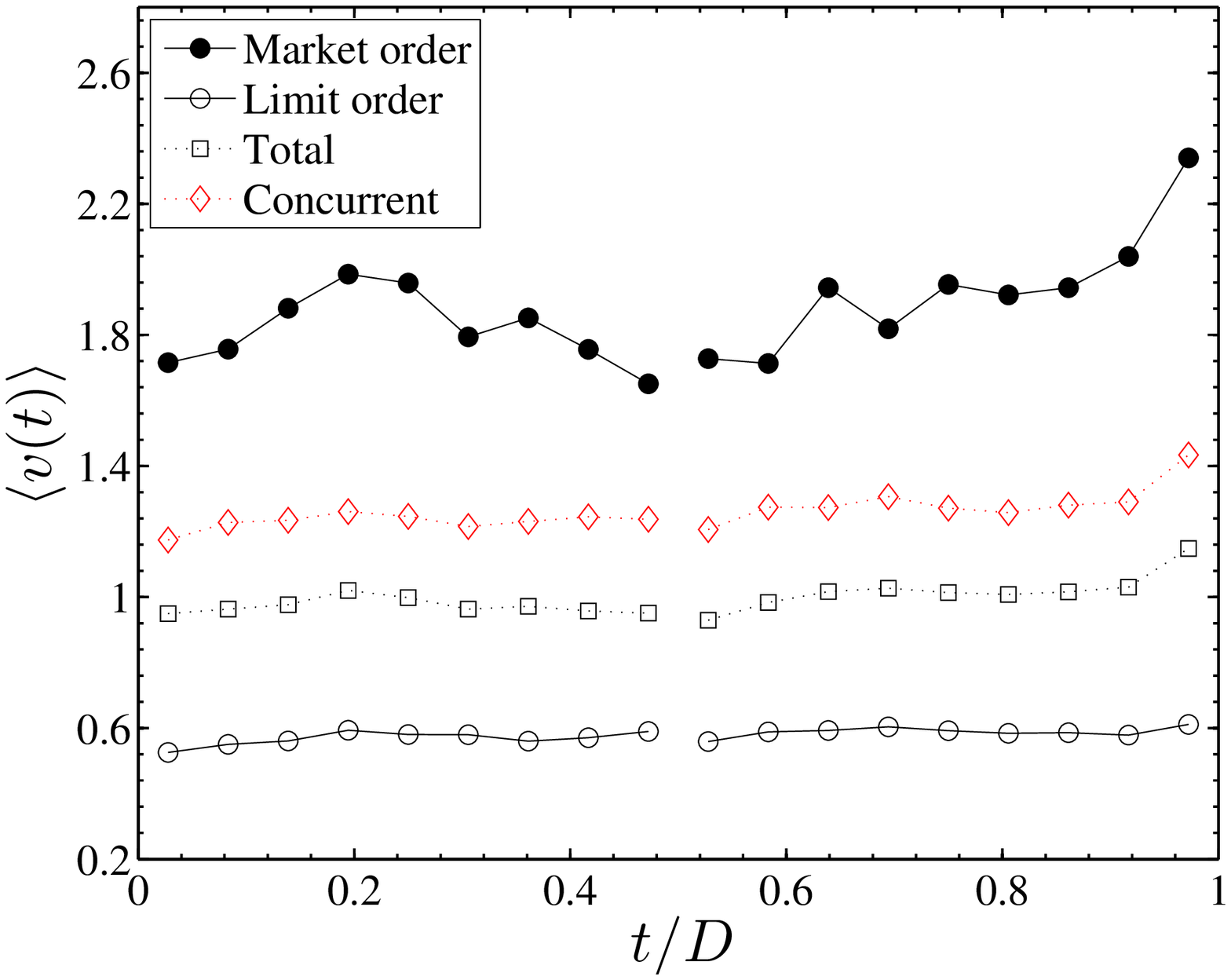}}\label{Profile_part_b}}%
\\
\subfigure[]{
\resizebox*{5cm}{!}{\includegraphics{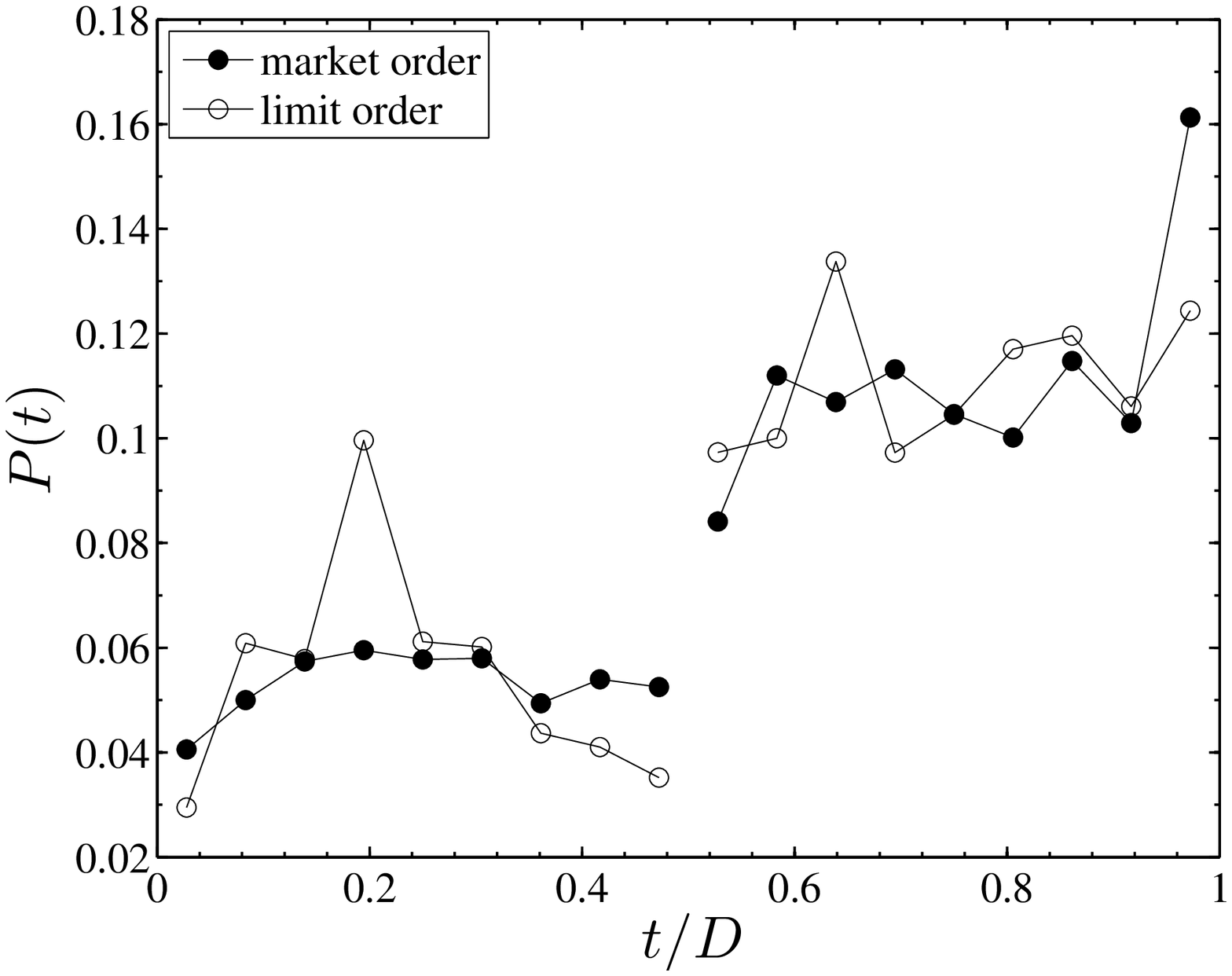}}\label{Profile_part_c}}%
\subfigure[]{
\resizebox*{5cm}{!}{\includegraphics{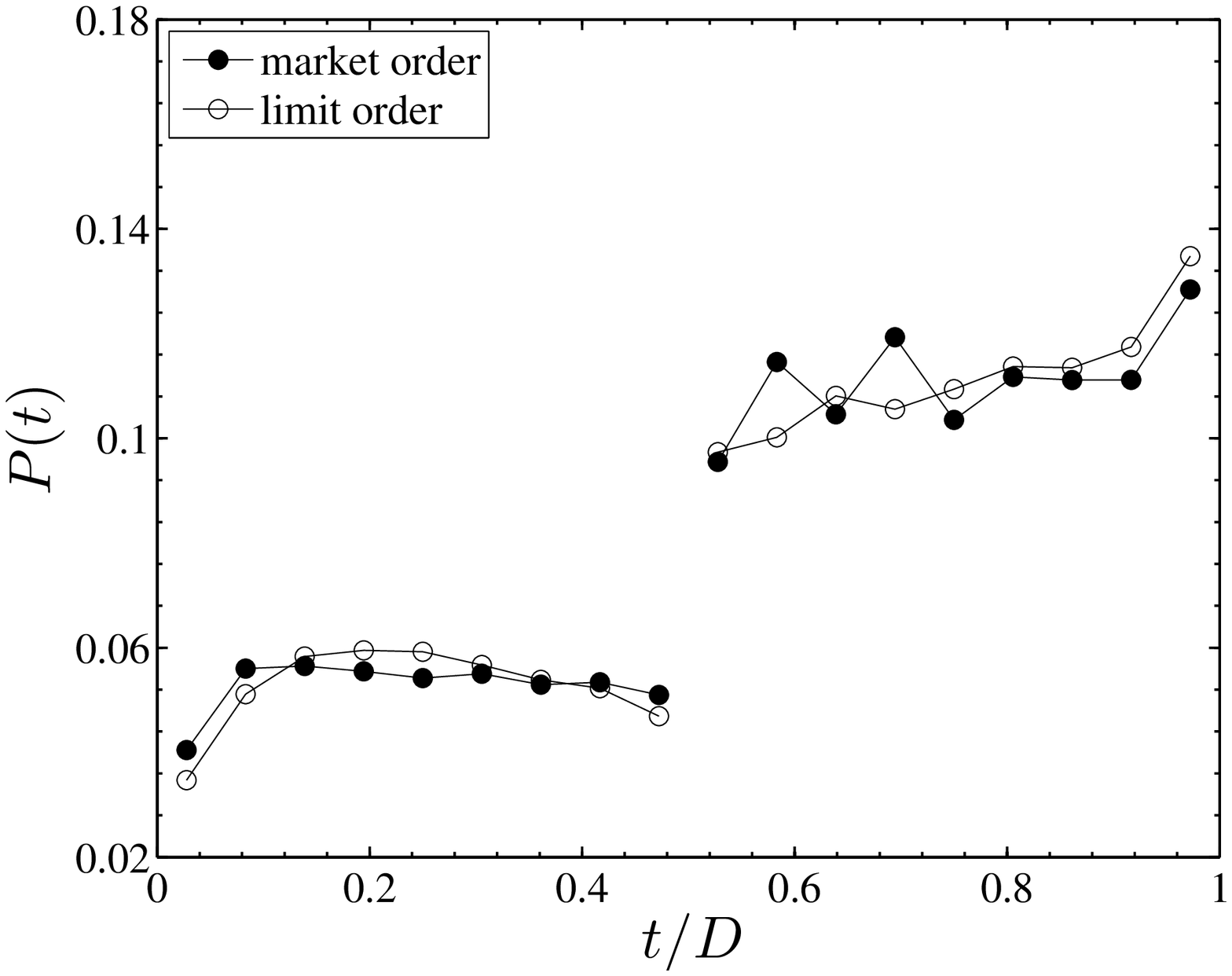}}\label{Profile_part_d}}%
\\
\subfigure[]{
\resizebox*{5cm}{!}{\includegraphics{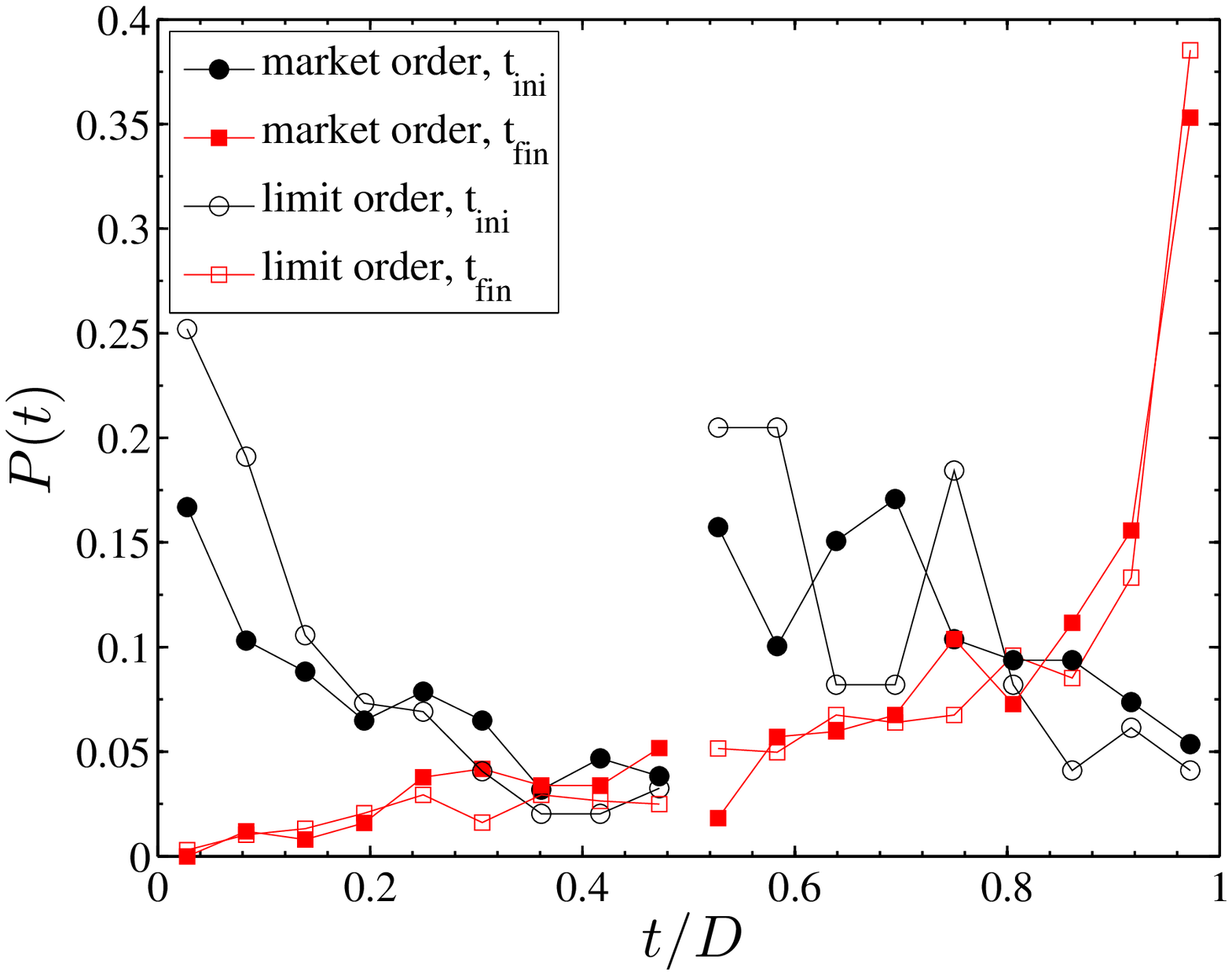}}\label{Profile_part_e}}%
\subfigure[]{
\resizebox*{5cm}{!}{\includegraphics{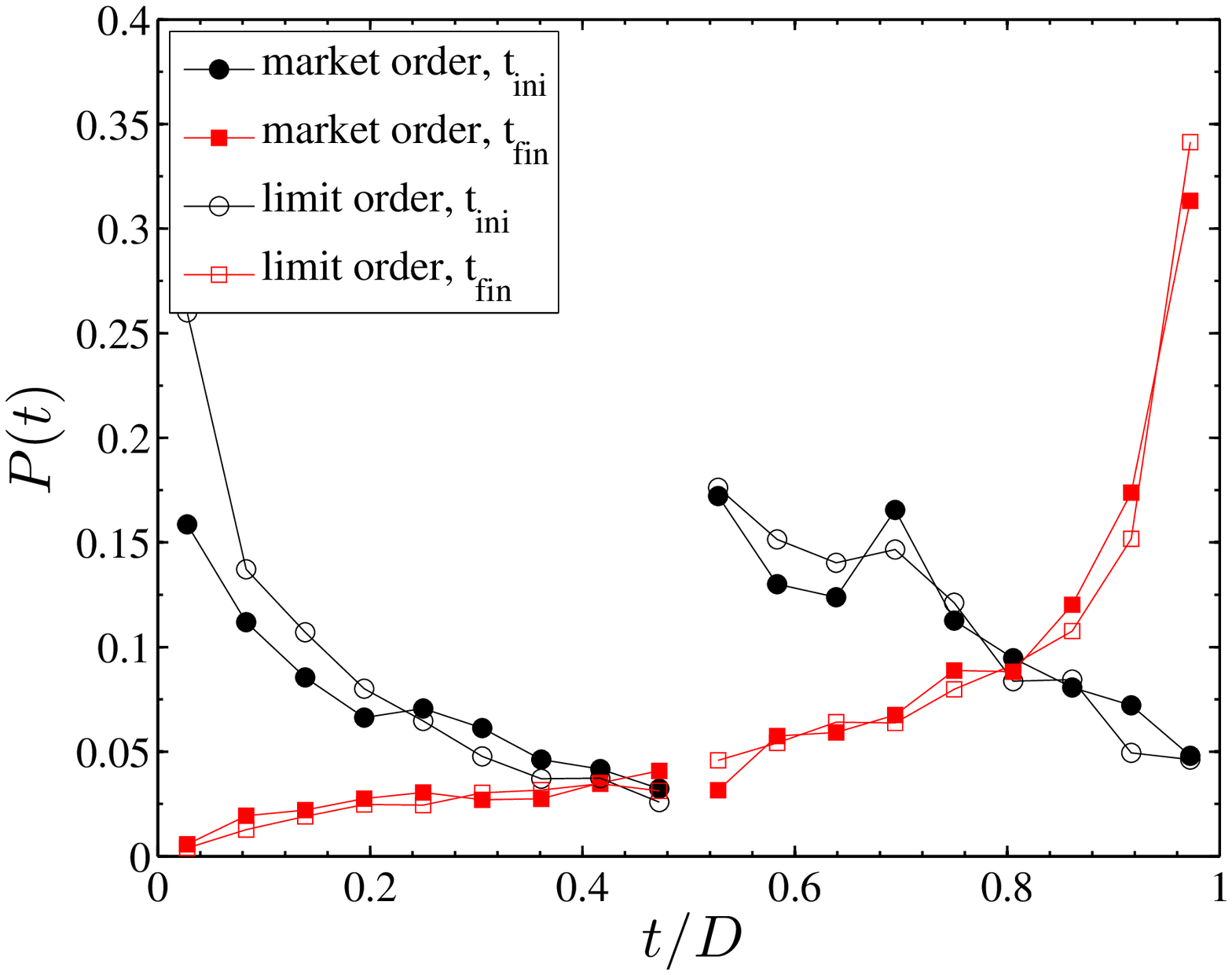}}\label{Profile_part_f}}%
\\
\subfigure[]{
\resizebox*{5cm}{!}{\includegraphics{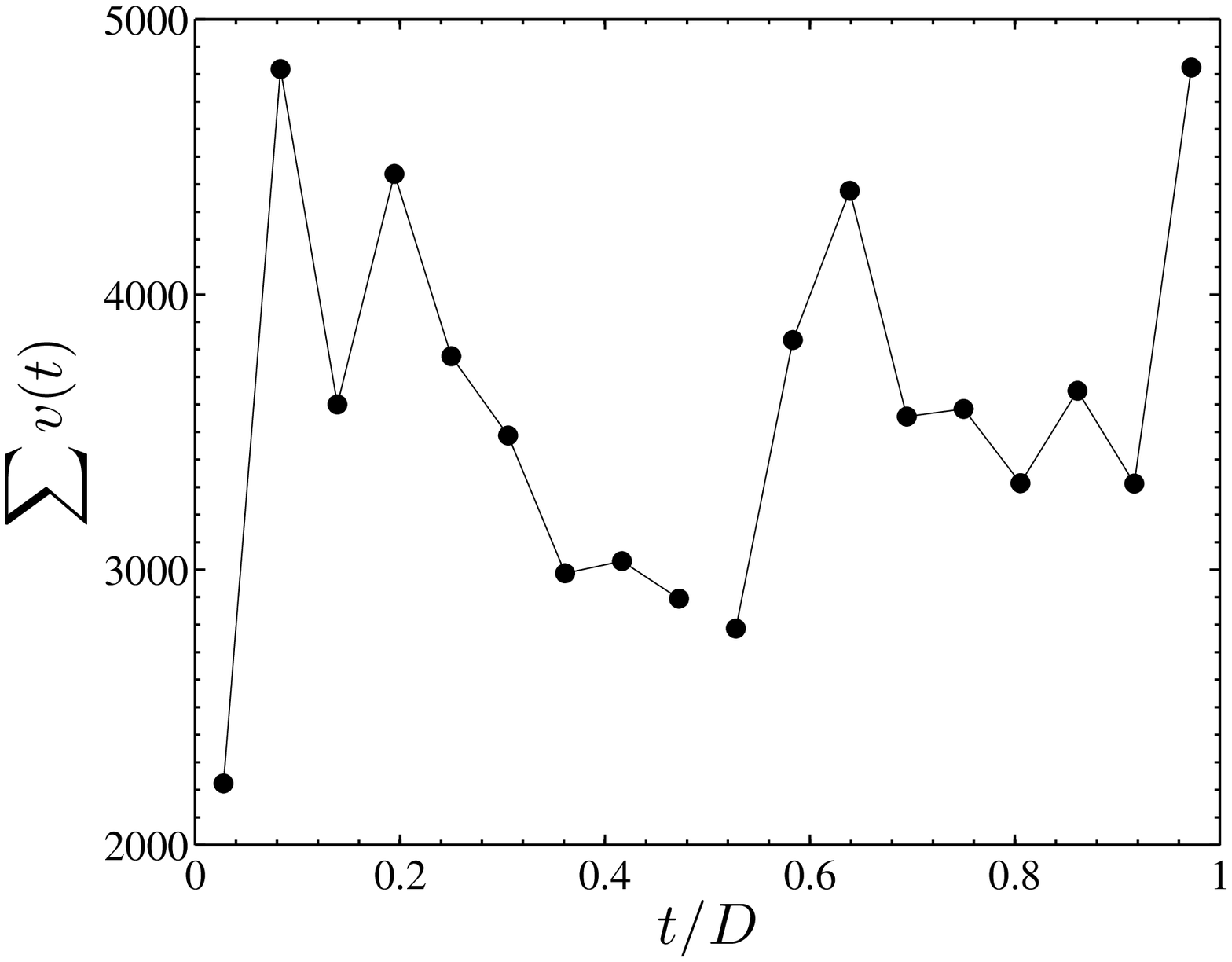}}\label{Profile_part_g}}%
\subfigure[]{
\resizebox*{5cm}{!}{\includegraphics{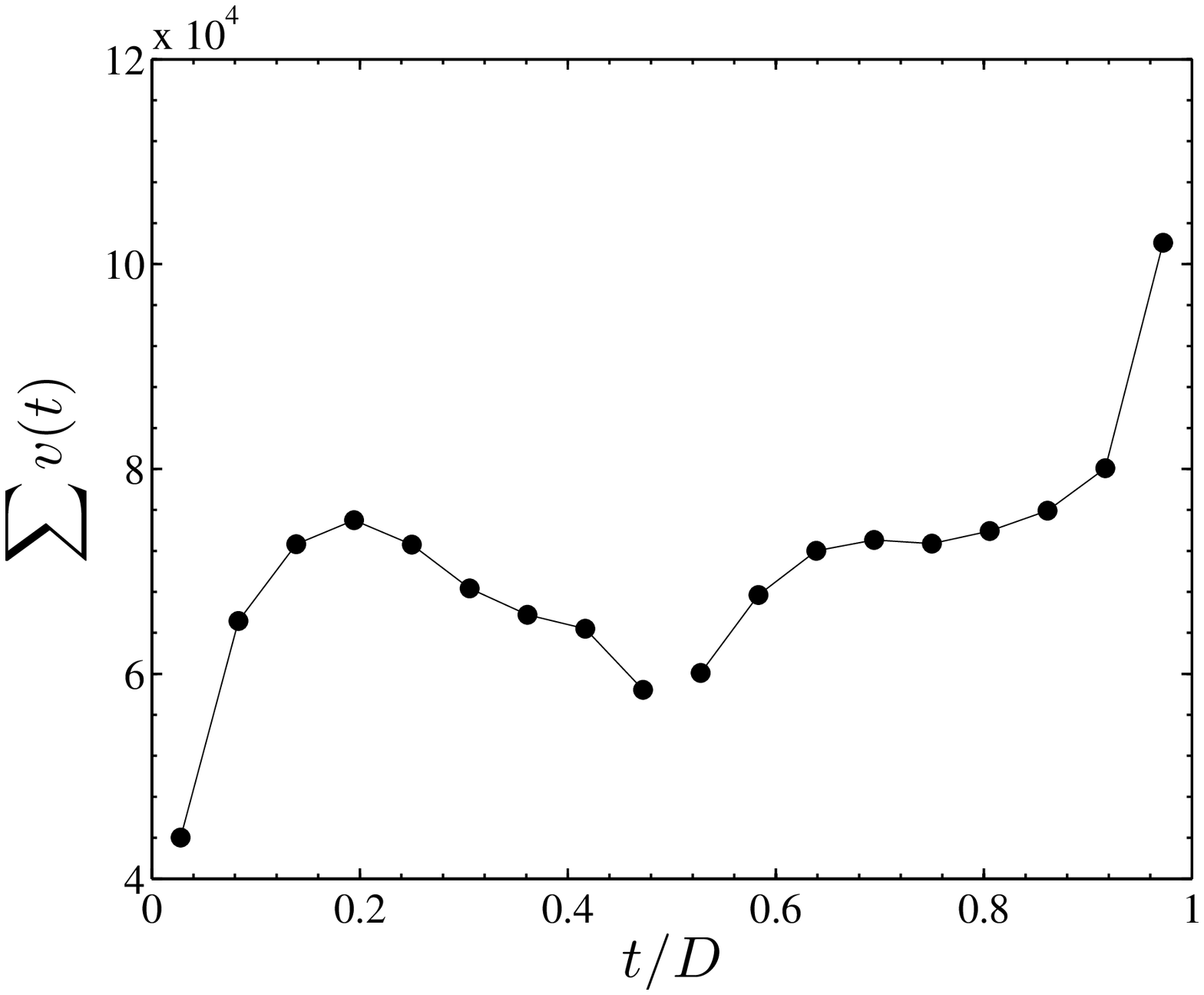}}\label{Profile_part_h}}%
\label{Profile}
\end{minipage}
\end{center}
\caption{(Color online) Trading profile of trade packages finished within one day: mean trading volume $\langle v(t) \rangle$ for (a) institutions and (b) individuals, probability distributions of the transaction time $t$ for (c) institutions and (d) individuals, probability distributions of the initial time $t_{ini}$ and final time $t_{fin}$ for (e) institutions and (f) individuals, and total trading volume $\sum v(t)$ for (g) institutions and (h) individuals. The time $t$ is measured with respect to the time of a day.}
\end{figure}

In Figures~\ref{Profile_part_a} and ~\ref{Profile_part_b}, the mean trading volume $\langle v(t) \rangle$ of individual transactions inside trade packages is plotted as a function of the normalized time $t/D$ for institutions and individuals respectively. Since the mean trading volume may differ for the transactions with different aggressiveness, we calculate $\langle v(t) \rangle$ of the transactions executed as market orders (filled black circles) and limit orders (empty black circles). One observes that institutions prefer to place large market orders close to the opening time, while individuals are more likely to place large market orders close to the closing time of the day. Moreover, $\langle v(t) \rangle$ of market orders is much more significant than that of limit orders, which indicates that the trade packages are accomplished mostly by market orders. Therefore, $\langle v(t) \rangle$ of the total market orders and limit orders (black squares) shows a similar profile to that of the market orders. The mean trading volume of the transactions concurrently traded with trade packages (red diamonds) also has a similar profile, slightly larger than that for the total market orders and limit orders, different from that observed in BME and LSE markets \citep{Clauset-Shalizi-Newman-2009-SIAMR}. These results are quite robust for trade packages of both institutions and individuals.

To complectly understand the trading profile, we further consider the number of transactions executed at time $t$. We calculate the probability distribution $P(t)$ of the transaction time $t$, defined as the proportion between number of transactions traded at time $t$ to the total number of transactions inside trade packages. As shown in Figure~\ref{Profile_part_c}, $P(t)$ for institutions is relatively small close to the opening time of the day. It may be explained by the phenomena that the mean volume of individual transactions is significant large then, and only a few numbers of transactions could accomplish the purchase or sale of large amounts of shares. There is a midday break during the lunch time in Chinese stock market, and the surge of $P(t)$ soon after midday break may be caused by the entrance of many institutions who start trade packages in the afternoon. An increase of $P(t)$ is also observed close to the closing time of the day, this may because that all the trade packages within one day should be accomplished at that time. The probability distribution $P(t)$ for individuals shows a similar profile to that for institutions. We also measure the probability distribution of the initial and final time of trade packages. As shown in Figures~\ref{Profile_part_e} and ~\ref{Profile_part_f}, for both institutions and individuals a significant fraction of trade packages are started at the opening time and finished at the closing time. Moreover, there are quite a number of trade packages start after the midday break, and this may explain the surge of $P(t)$ soon after the midday break.

We finally investigate the total trading volume of individual transactions inside trade packages traded at time $t$, i.e., the product of the mean volume of individual transactions and the number of transactions. In Figures~\ref{Profile_part_g} and ~\ref{Profile_part_h}, the total trading volume $\sum v(t)$ of transactions inside trade packages is plotted with respect to time $t$ for institutions and individuals respectively. For the opening hours in the morning, $\sum v(t)$ for institutions shows a maximum at around $0.1$ day (approximately at 10:00), about half hour earlier than the location of the maximum for individuals. After the midday break, $\sum v(t)$ for institutions shows a rapid increase and exhibits a maximum at round $0.65$ day (approximately at 13:30), while no such peak is observed for individuals. Large $\sum v(t)$ is further observed close to the closing time for both institutions and individuals. The fact that institutions start their trade packages earlier than individuals implies institutions may be more informed than individuals.

\section{Price impact}
\label{sec:Impact}

\subsection{Price impact of trade packages}
\label{sec:PackageImpact}

The price impact is a very important issue in financial studies, and a large number of studies have focused attention on this topic. In this section,
we first consider the price impact of trade packages, generally measured as the difference between the price prior and after the trade package \citep{Chan-Lakonishok-1995-JF,Gallagher-Looi-2006-AF}. For a trade package accomplished between time $t$ and $t+T$, we follow the work by \citet{Moro-Vicente-Moyano-Gerig-Farmer-Vaglica-Lillo-2009-PRE}, and define the price impact as the difference between the logarithmic price of the first and last transaction of the trade package
\begin{equation}
   r = \ln p(t+T) - \ln p(t).
   \label{Eq:package:return}
\end{equation}
The scaled price impact is obtained by taking into account the normalization condition and the sign of the trade package
\begin{equation}
   R = s r/\langle |r| \rangle,
   \label{Eq:package:impact}
\end{equation}
where $s=+1$ ($-1$) represents the package with mostly buy (sell) trades, and $r$ is normalized by the mean absolute return $\langle |r| \rangle$ of the relevant stock.

\subsubsection{$R$ vs $T$}

\begin{figure}
\begin{center}
\begin{minipage}{100mm}
\subfigure[]{
\resizebox*{5cm}{!}{\includegraphics{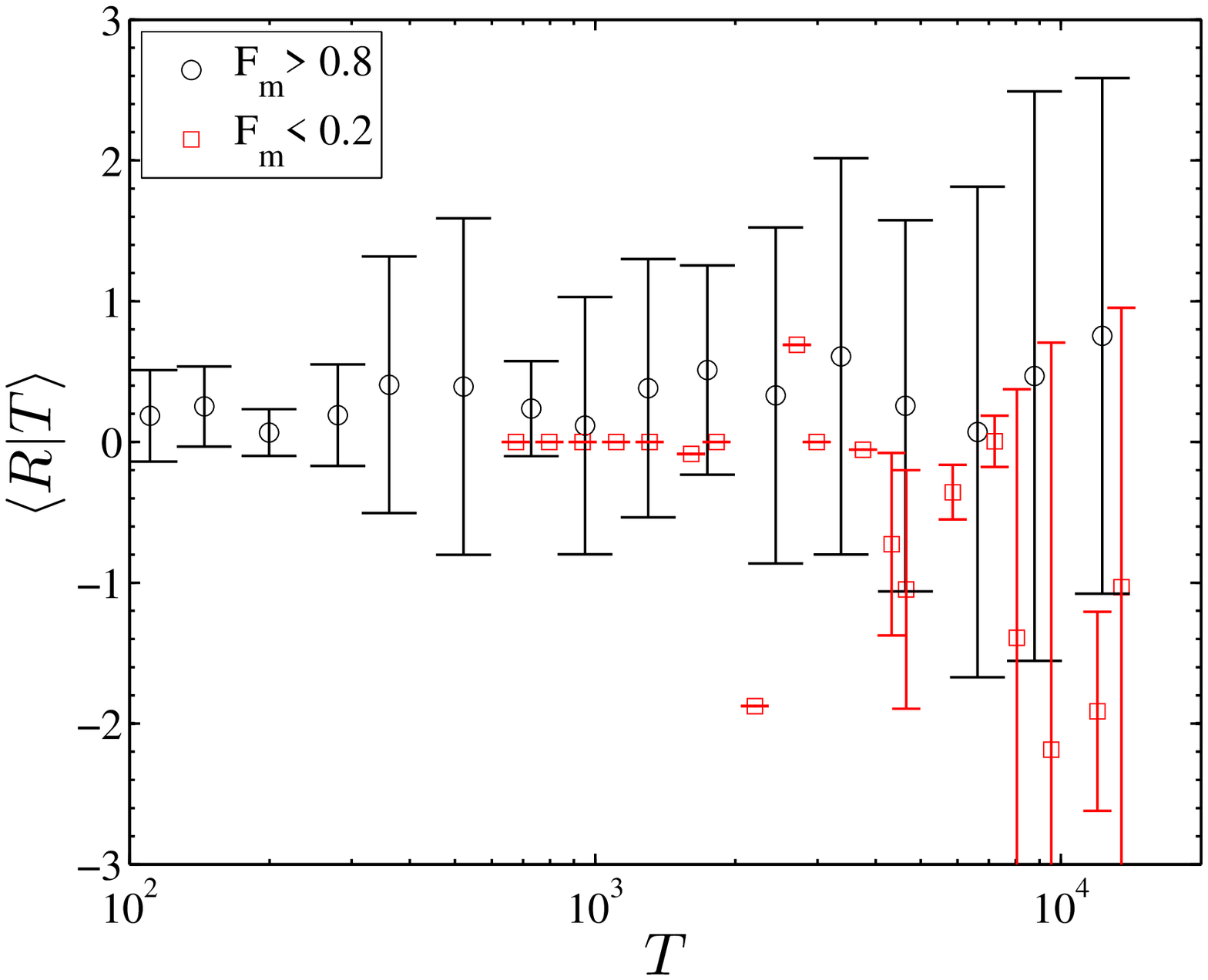}}\label{Impact-package_part_a}}%
\subfigure[]{
\resizebox*{5cm}{!}{\includegraphics{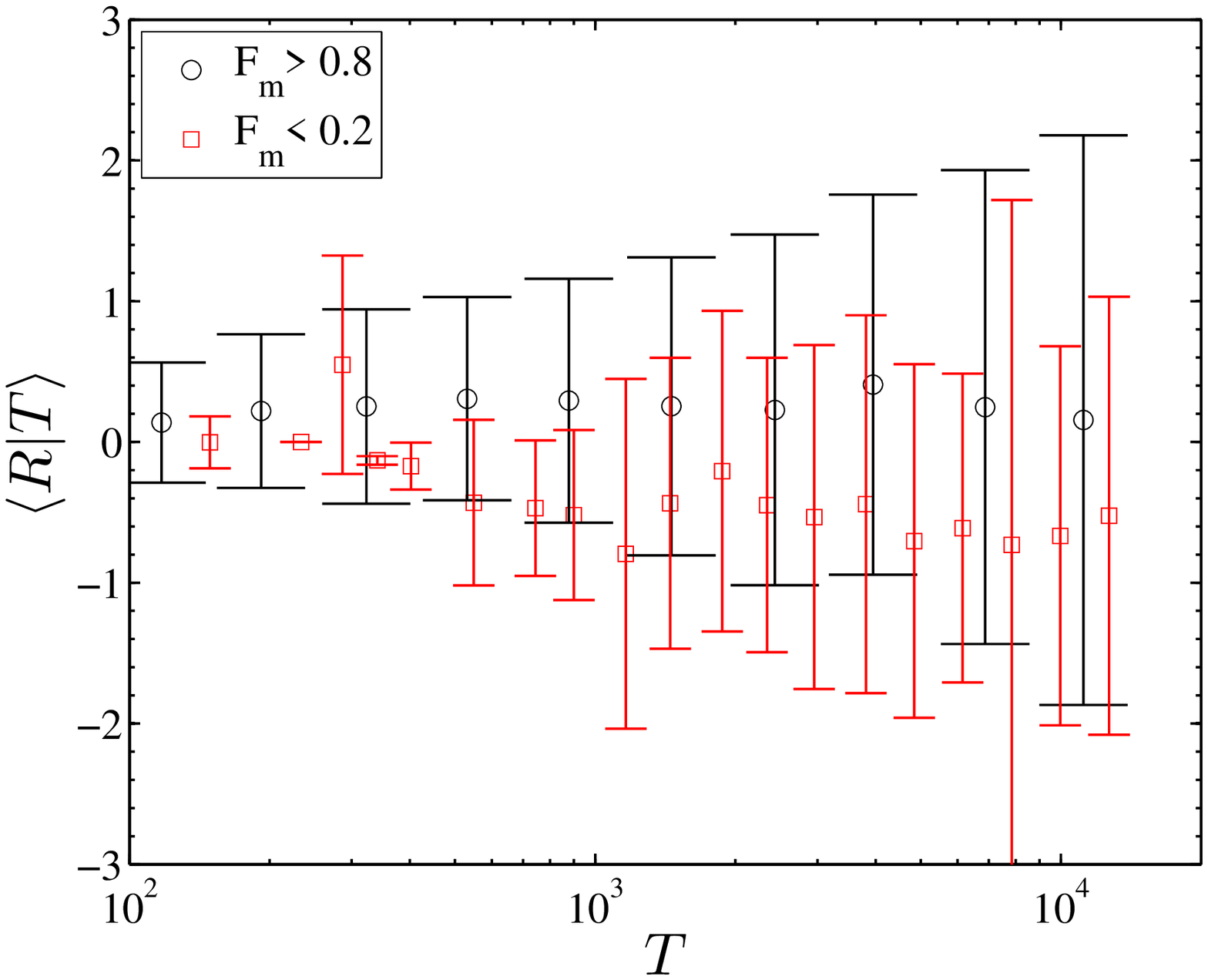}}\label{Impact-package_part_b}}%
\\
\subfigure[]{
\resizebox*{5cm}{!}{\includegraphics{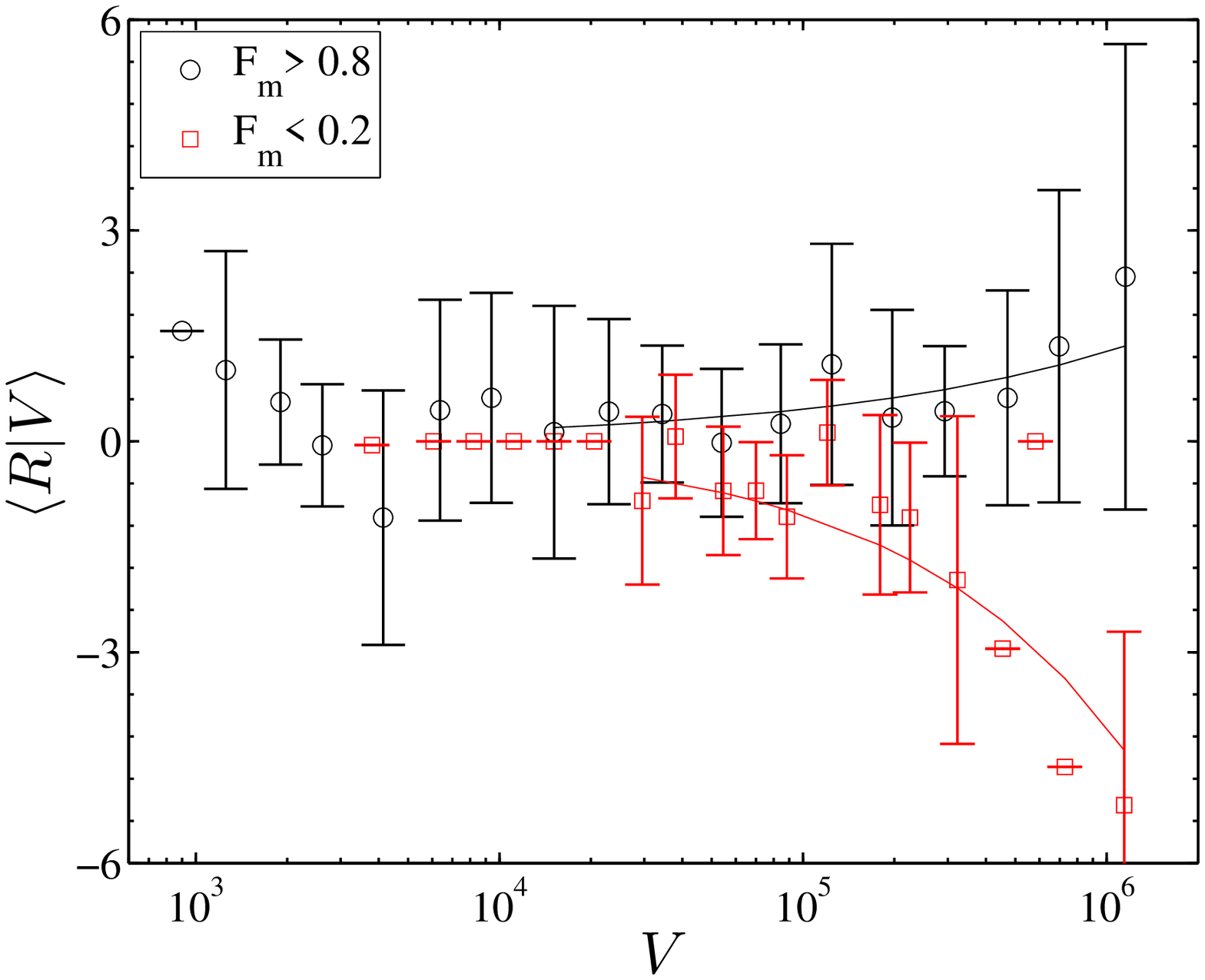}}\label{Impact-package_part_c}}%
\subfigure[]{
\resizebox*{5cm}{!}{\includegraphics{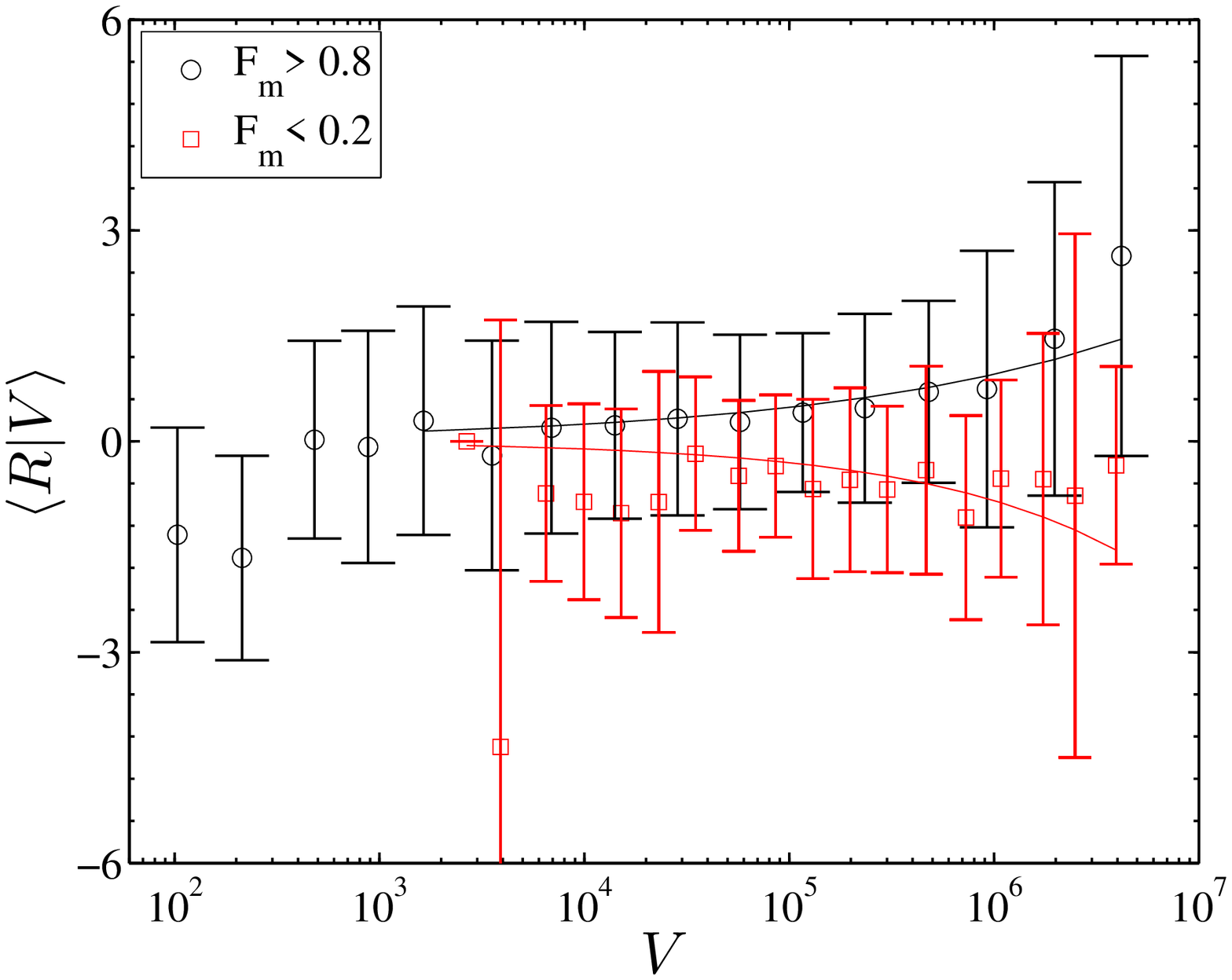}}\label{Impact-package_part_d}}%
\label{Impact-package}
\end{minipage}
\end{center}
\caption{(Color online) Mean conditional scaled price impact of trade packages finished within one day: $\langle R|T \rangle$ for (a) institutions and (b)individuals, and $\langle R|V \rangle$ for (c) institutions and (d) individuals. The solid curves are power-law fits with exponents depicted in Table~\ref{TB:RV}.}
\end{figure}

We investigate the dependence of the scaled price impact $R$ on the execution time $T$. We only consider the trade packages finished within one day, since the price impact persists over more than one day not only capture the information contained in the trade package but also be affected by the nonsuccessive trades overnight. In Figures~\ref{Impact-package_part_a} and ~\ref{Impact-package_part_b}, the mean conditional scaled price impact $\langle R|T \rangle$ is plotted as a function of the execution time $T$ for institutions and individuals respectively. A trade package can be accomplished with different component of market orders and limit orders. To identify it, we calculate the variable $F_m$ defined as the fraction of volumes done through market orders within trade packages. The mean conditional price impact $\langle R|T \rangle$ for trade packages with large and small fraction of market orders, i.e., $F_m>0.8$ and $F_m<0.2$, are represented separately in the figure.

Despite of strong fluctuation in the conditional price impact, the mean conditional price impact $\langle R|T \rangle$ with $F_m>0.8$ is mostly positive over the whole range of $T$ for both institutions and individuals. This indicates that the trade package has a nonnegligible price impact over the whole range of execution time up to one day. The cumulative effect of a sequence of mostly buyer-initiated trades is more like to raise the price, while the cumulative effect of a sequence of mostly seller-initiated trades is more likely to reduce the price. Moreover, $\langle R|T \rangle$ with $F_m<0.2$ is mostly negative, and this implies that the price impact of a trade packages is not dominated by the sign of the most trades when there is a low fraction of market orders. We further perform the analysis of variance (ANOVA) to test if the means of the conditional price impact for different $T$ are equal. For institutions, we obtain a $p$-value 0.919 for trade packages with $F_m>0.8$, and a $p$-value 0.787 for trade packages with $F_m<0.2$. Similar large $p$-values are observed for trade packages of individuals, which indicates that $\langle R|T \rangle$ does not have much difference for various $T$. No clear tendency of $\langle R|T \rangle$ is observed in our study.

\subsubsection{$R$ vs $V$}

We then investigate the dependence of the scaled price impact $R$ on the total trading volume $V$. The mean conditional scaled price impact $\langle R|V \rangle$ with $F_m>0.8$ and $F_m<0.2$ for both institutions and individuals are plotted as a function of the total trading volume $V$ in Figures~\ref{Impact-package_part_c} and ~\ref{Impact-package_part_d}. For both institutions and individuals, $\langle R|V \rangle$ is mostly positive for trade packages with $F_m>0.8$, while mostly negative for trade packages with $F_m<0.2$. This further confirms our previous finding that the cumulative impact of a trade package is dominated by the sign of the most trades when there is a large fraction of market orders. We also use the ANOVA to compare the means of the conditional price impact for different $V$, and find very small $p$-values for trade packages of both institutions and individuals with $F_m>0.8$ and $F_m<0.2$, indicating the means differ with the variation of $V$. Similar to the price-volume relation revealed in many empirical studies, we assume the absolute $\langle R|V \rangle$ for large $V$ follows a power law
\begin{equation}
  | \langle R|V \rangle |=A V^{\gamma}.
  \label{Eq:PL:RV}
\end{equation}
The power-law fits with estimated parameters $A$ and $\gamma$ listed in Table~\ref{TB:RV} are illustrated in Figures~\ref{Impact-package_part_c} and ~\ref{Impact-package_part_d}. For trade packages with $F_m>0.8$, $\gamma$ is about 0.447 for institutions, and 0.295 for individuals. The exponent $\gamma$ for trade packages with $F_m<0.2$ is slightly larger. The power-law increase of $\langle R|V \rangle$ at large scales of $V$ may correspond to the purchase or sale of large amounts of shares with the motive to adjust share inventory. However, $\langle R|V \rangle$ for trade packages of institutions with $F_m>0.8$ does not increase monotonously with the increase of $V$. The variable $\langle R|V \rangle$ shows a decreasing tendency with increasing $V$ when $V$ is small. This indicates that small trading volumes may also cause large price changes, and it may refer to the purchase or sale of small amounts of shares for the purpose of stock price adjustment.

\begin{table}
\begin{center}
\begin{minipage}{80mm}
  \tbl{Estimated parameters for the power-law fit of $\langle R|V \rangle$ for trade packages of both institutions and individuals.}
{\begin{tabular}{cccccc}
  \toprule
    \multirow{3}*[2mm]  & \multicolumn{2}{c}{Institution} & & \multicolumn{2}{c}{Individual}\\  %
  \cline{2-3} \cline{5-6}
    & $A$ & $\gamma$ & & $A$ & $\gamma$ \\
  \colrule
   $F_m > 0.8$ &   $0.003$ & $0.447(0.223)$ & & $0.016$ & $0.295(0.149)$ \\%
  \colrule
   $F_m < 0.2$ &   $0.001$ & $0.590(0.130)$ & & $0.002$ & $0.444(1.290)$ \\%
  \botrule
  \end{tabular}}
\label{TB:RV}
\end{minipage}
\end{center}
\end{table}

\subsection{Instantaneous price impact of transactions inside trade packages}
\label{sec:Inst}

We have shown the price impact of the entire trade packages, and then we investigate the price impact of individual transactions inside trade packages. Given a transaction $i$ inside trade package traded at time $t$, $p(t_-)$ is the price before this transaction, and $p(t)$ is the resultant price immediate after it. The price impact of the transaction $i$ inside trade package is defined as
\begin{equation}
   r_i = \ln p(t) - \ln p(t_-),
   \label{Eq:return}
\end{equation}
which is the instantaneous change of logarithmic price contributed from the transaction $i$. To make the price impact comparable over different stocks, the price impact of a particular stock is normalized by its mean absolute value
\begin{equation}
   R_i = s_i r_i/\langle |r_i| \rangle,
   \label{Eq:impact}
\end{equation}
where $s_i=+1$ for buy trades and $s_i=-1$ for sell trades.

\subsubsection{$R_i$ vs $t$}

\begin{figure}
\begin{center}
\begin{minipage}{100mm}
\subfigure[]{
\resizebox*{5cm}{!}{\includegraphics{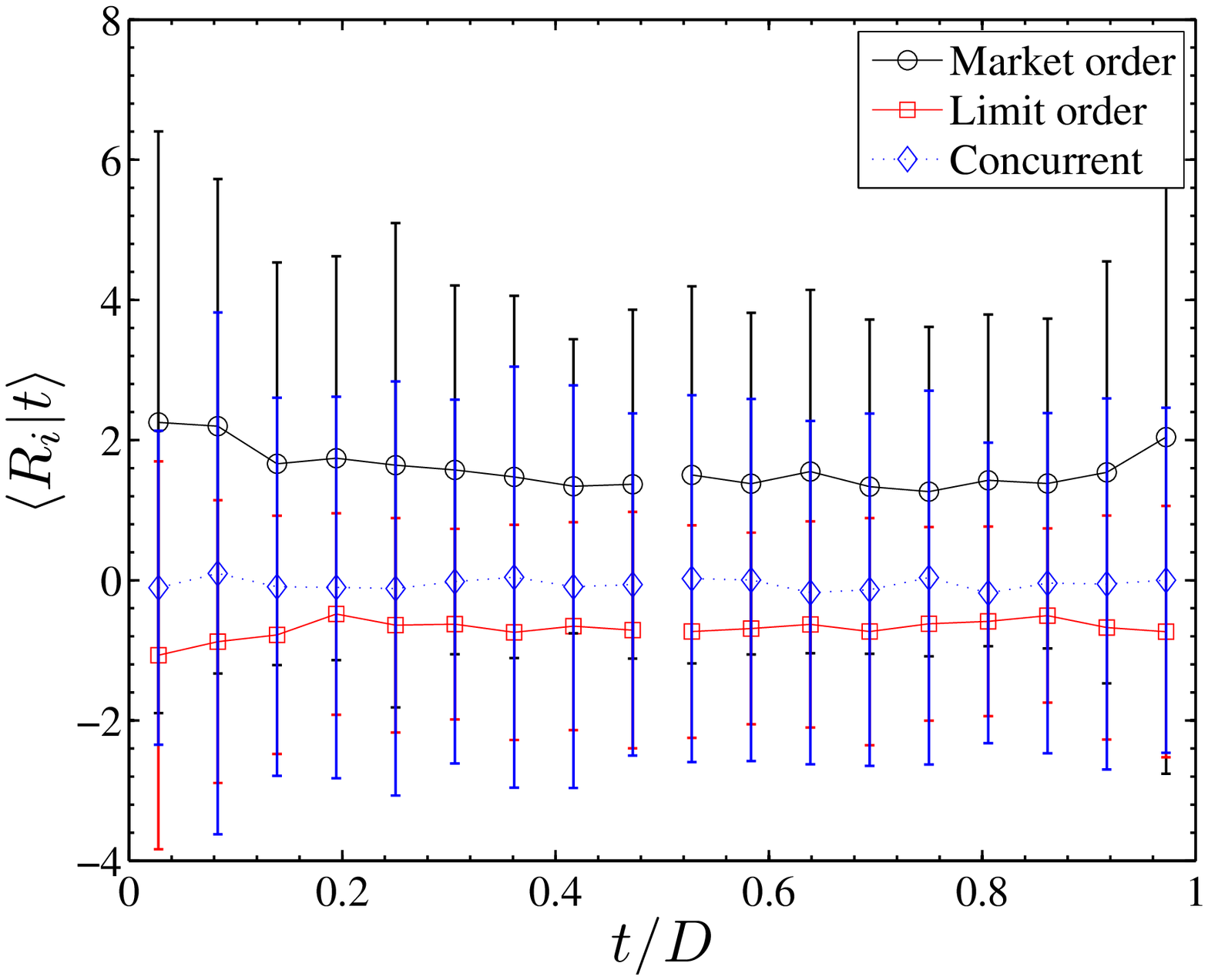}}\label{Impact-transaction_part_a}}%
\subfigure[]{
\resizebox*{5cm}{!}{\includegraphics{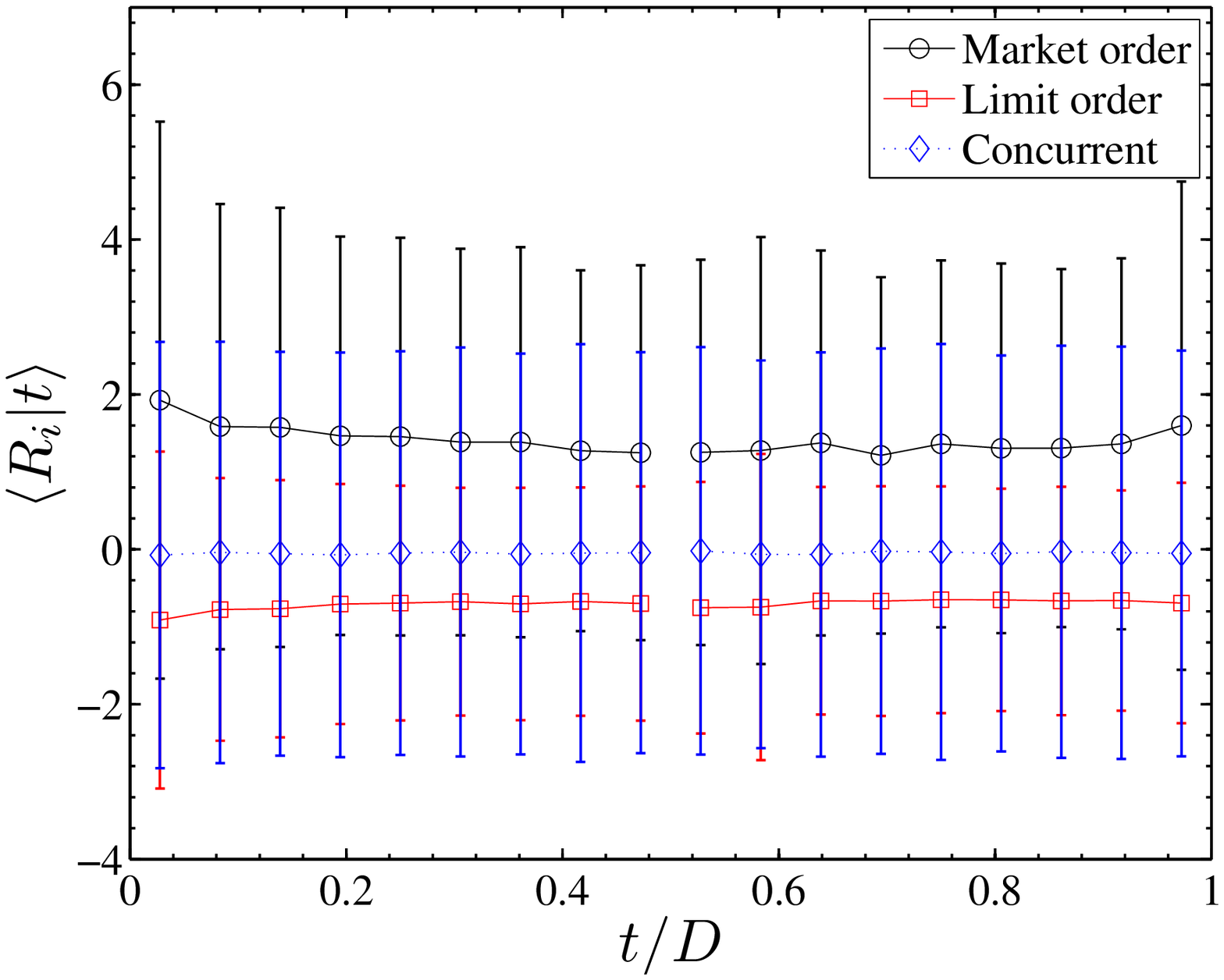}}\label{Impact-transaction_part_b}}%
\\
\subfigure[]{
\resizebox*{5cm}{!}{\includegraphics{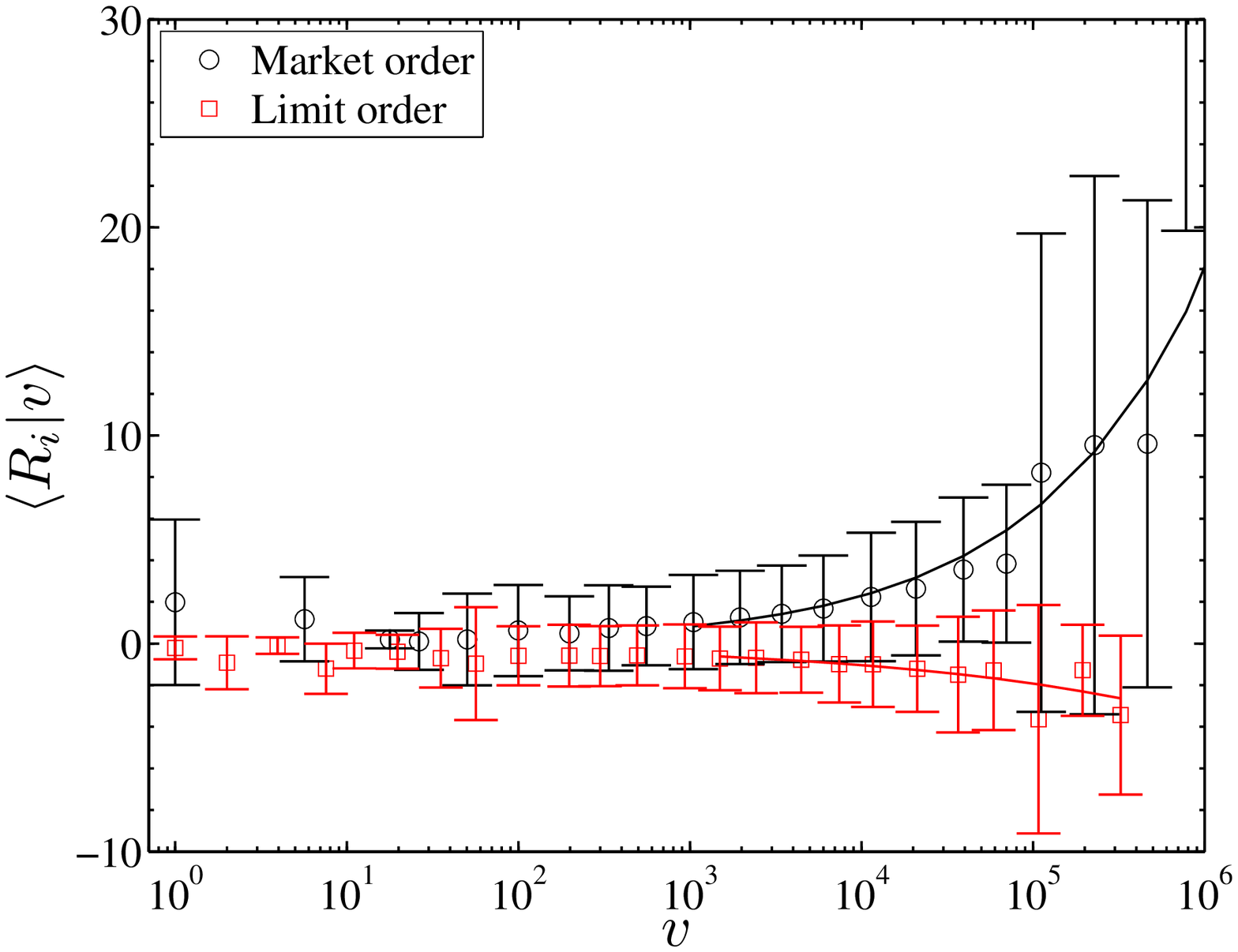}}\label{Impact-transaction_part_c}}%
\subfigure[]{
\resizebox*{5cm}{!}{\includegraphics{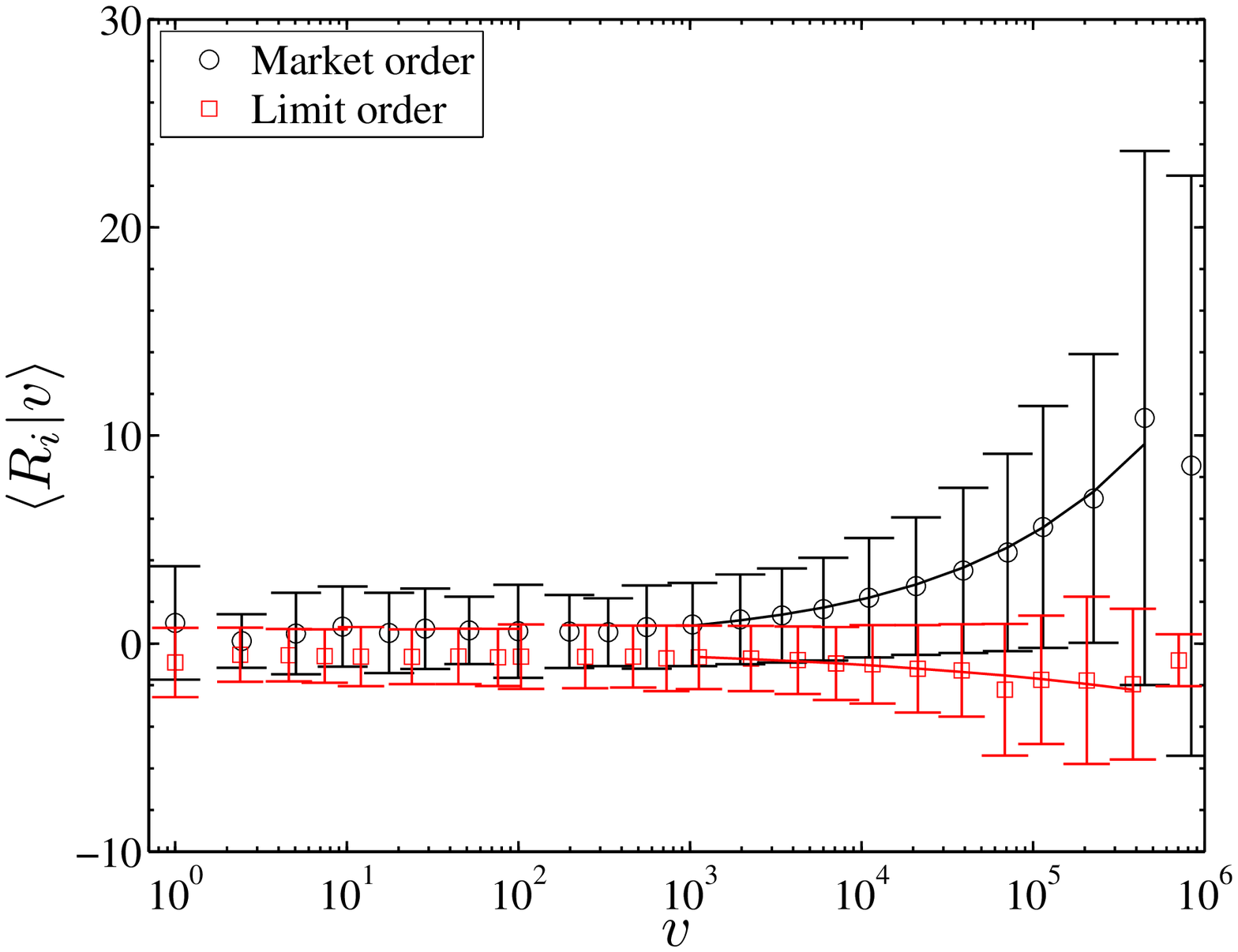}}\label{Impact-transaction_part_d}}%
\label{Impact-transaction}
\end{minipage}
\end{center}
\caption{(Color online) Mean conditional scaled price impact of individual transactions inside trade packages finished within one day: $\langle R_i|t \rangle$ for (a) institutions and (b) individuals, and $\langle R_i|v \rangle$ for (c) institutions and (d) individuals. The time $t$ is measured with respect to the time of a day.}
\end{figure}

We measure the mean conditional scaled price impact $\langle R_i|t \rangle$ of individual transactions conditioned on the time $t$. In Figures~\ref{Impact-transaction_part_a} and ~\ref{Impact-transaction_part_b}, $\langle R_i|t \rangle$ is plotted as a function of the time $t$ for institutions and individuals respectively. The mean conditional scaled price impact of the transactions executed as market orders and limit orders are represented by black circles and red squares in the figure. For market orders, $\langle R_i|t \rangle$ displays a concave U-shaped profile. Though the error bar is quite large, $\langle R_i|t \rangle$ for market orders is mostly positive.  For limit orders, $\langle R_i|t \rangle$ is mostly negative, and displays an inverted U-shaped profile. The sign of $\langle R_i|t \rangle$ indicates that the direction of the price change is driven by the market order. In other words, the buyer-initiated trades raise the price, and the seller-initiated trades reduce the price. Moreover, the U-shaped and inverted U-shaped profiles imply that the price impact close to the opening and closing time is larger than the price impact during the remainder of the day. We also investigate the price impact of the transactions concurrently traded with those transactions inside trade packages
\begin{equation}
   R_{con} = s_i r_{con}/\langle |r_i| \rangle,
   \label{Eq:impact:con}
\end{equation}
where $s_i$ and $\langle |r_i| \rangle$ are the sign and the mean absolute return of transactions inside trade packages, and $r_{con}$ represents the price return caused by the transactions concurrently traded with those transactions inside trade packages. The mean conditional scaled price impact $\langle R_{con}|t \rangle$ (blue diamonds) fluctuates around zero, which indicates that the price impact of market trades is not merely driven by the consecutive trades by large investors.

\subsubsection{$R_i$ vs $v$}

To investigate the dependence of the price impact of transactions inside trade packages on their trading volumes, we calculate the mean conditional scaled rice impact $\langle R_i|v \rangle$. In Figures~\ref{Impact-transaction_part_c} and ~\ref{Impact-transaction_part_d}, $\langle R_i|v \rangle$ is plotted for institutions and individuals respectively. One observes that $\langle R_i|v \rangle$ for transactions executed as market orders is mostly positive while that for transactions executed as limit orders is mostly negative. This further confirms the result that the market order determines the direction of the price change as observed in $\langle R_i|t \rangle$. Similar to $\langle R|V \rangle$ for trade packages, the absolute $\langle R_i|v \rangle$ of transactions inside trade packages follows a power law
\begin{equation}
  |\langle R_i|v \rangle|=B v^{k},
  \label{Eq:PL:Rv}
\end{equation}
when $v$ is large enough $v>10^3$. The estimated parameters $B$ and $k$ for the power-law fit of $\langle R_i|v \rangle$ are listed in Table~\ref{TB:Rv}. The exponent $k$ for transactions executed as market orders is larger than that for transactions executed as limit orders, which indicates that for the isolated transactions inside trade packages the price impact is remarkably affected by the market order. However, the cumulative impact of trade packages with small fraction of market orders is comparable with the cumulative impact of trade packages with large fraction of market orders. In fact, the exponent $\gamma$ of $\langle R|V \rangle$ for packages with $F_m<0.2$ is slightly larger than that for packages with $F_m>0.8$. This may because the investors use different strategies for the accomplishment of trade packages. Like a sequence of mostly buyer-initiated trades, a sequence of sales mostly executed as limit orders can also raise the price, affected by the buyer-initiated trades on the opposite sides.

\begin{table}
\begin{center}
\begin{minipage}{80mm}
  \tbl{Estimated parameters for the power-law fits of $\langle R_i|v \rangle$ for transactions inside trade packages of both institutions and individuals.}
{\begin{tabular}{cccccc}
  \toprule
   \multirow{3}*[2mm]  & \multicolumn{2}{c}{Institution} & & \multicolumn{2}{c}{Individual}\\  %
  \cline{2-3} \cline{5-6}
    & $B$ & $k$ & & $B$ & $k$ \\
  \colrule
   market order &  $0.04$ & $0.45(0.11)$ & & $0.05$ & $0.40(0.02)$ \\%
  \colrule
   limit order  &  $0.09$ & $0.27(0.12)$ & & $0.15$ & $0.21(0.05)$ \\%
  \botrule
  \end{tabular}}
\label{TB:Rv}
\end{minipage}
\end{center}
\end{table}

\subsection{Temporary price impact of transactions inside trade packages}
\label{sec:Temp}

\subsubsection{Impact of trading volumes on current and following price returns}

Numerous studies have focused on the influence of trading volumes or volume imbalances on price returns \citep {Wood-McInish-Ord-1985-JF,Karpoff-1987-JFQA,Gallant-Rossi-Tauchen-1992-RFS,Saatcioglu-Starks-1998-IJF,Lillo-Farmer-Mantegna-2003-Nature,Plerou-Gopikrishnan-Gabaix-Stanley-2002-PRE,Chordia-Subrahmanyam-2004-JFE}. We analyze the temporary impact of transactions inside trade packages in relation to their trading volumes. Suppose $r(t)=\ln p(t) - \ln p(t-1)$ is the price return at time $t$, $v(t)$ is the trading volume of a transaction inside trade package traded at time $t$. The returns of a certain stock are normalized by its standard deviation. To explain the influence of the trading volume at time $t$ on the price return at later time $t+i$, we estimate a regression of the following form
\begin{equation}
   R(t+i) = \beta_0^{\ast} + \beta_i s \ln v(t)+\epsilon(t),
   \label{Eq:impact:tem1}
\end{equation}
where $s$ is the sign of the transaction. We measure the impact of the trading volumes $v(t)$ within trade package on the current price return $i=0$ and the following price returns $i=5,10,15,20,25$ seconds. It has been empirically verified that the price return has a power-law dependence on the trading volume \citep{Hasbrouck-1991-JF,Plerou-Gopikrishnan-Gabaix-Stanley-2002-PRE,Chordia-Subrahmanyam-2004-JFE,Zhou-2011-QF}, also confirmed in the measure of $\langle R_i|v \rangle$. The first-order approximation of the logarithmic return is assumed, and a linear regression of the return against logarithmic trading volume is used to approximate their power-law relations.

Table~\ref{TB:impact:tem1:ins} reports the estimated coefficients of regressions using the records of transactions inside trade packages executed as market orders from institutions. Positive coefficient $\beta_{0sec}$ is observed in all the 23 stocks, which further confirms the finding that the sign of price return is determined by the sign of the market order as revealed in the measure of $\langle R_i|t \rangle$. The relevant $t$-statistics are also depicted in Table~\ref{TB:impact:tem1:ins}. The impact of trading volume on the current price return is significant for most of stocks except for stocks $000009,000012,000429,000550$. This indicates that the instantaneous impact of the trading volumes executed as market orders is positive and significant. We find the price tends to reverse soon after the transaction inside trade package, showing negative $\beta_i$ with $i>0$. This reminds us of the significant price reversal after large price changes \citep{Zawadowski-Kertesz-Andor-2004-PA,Mu-Zhou-Chen-Kertesz-2010-NJP}. The magnitudes of $\beta_i$ with $i>0$ are significantly smaller than $\beta_{0sec}$, and decrease with the increase of time lag $i$. Coefficient $\beta_{5sec}$ is negative and significant for more than half of the stocks, and negative and significant $\beta_{10sec}$ is also observed in almost half of the stocks. This implies that for almost half of the stocks the trading volumes of transactions inside trade packages have temporary negative effects on price returns for at least $5-10$ seconds. With rare exceptions, the stocks have nonsignificant coefficients $\beta_i$ for large time lag $i\geq 15$ seconds. For the trading volumes of transactions inside trade packages executed as limit orders, a temporary impact is also observed. Negative $\beta_{0sec}$ is obtained for most of the stocks, indicating the price return has a sign opposite to that of the trading volume. Positive $\beta_{5sec}$ and $\beta_{10sec}$ further confirm the price reversal soon after the transaction inside trade package.

\begin{table}
\begin{center}
\begin{minipage}[c]{\linewidth}
  \tbl{Coefficients and $t$-statistics for the linear fit of Equation~(\ref{Eq:impact:tem1}) using the records of transactions inside trade packages executed as market orders from institutions. Coefficients marked with $^{\ast}$ are statistically significant at 5\% level.}
{\begin{tabular}{C{0.9cm}r@{.}lR{0.9cm}r@{.}lR{0.9cm}r@{.}lR{0.9cm}r@{.}lR{0.9cm}r@{.}lR{0.9cm}r@{.}lR{0.9cm}}
  \toprule
   Code &  \multicolumn{2}{c}{$\beta_{0}$} &  \multicolumn{1}{c}{$t$-stat} &  \multicolumn{2}{c}{$\beta_{5}$} &  \multicolumn{1}{c}{$t$-stat} &  \multicolumn{2}{c}{$\beta_{10}$} &  \multicolumn{1}{c}{$t$-stat} &  \multicolumn{2}{c}{$\beta_{15}$} &  \multicolumn{1}{c}{$t$-stat} &  \multicolumn{2}{c}{$\beta_{20}$} &  \multicolumn{1}{c}{$t$-stat} &  \multicolumn{2}{c}{$\beta_{25}$} &  \multicolumn{1}{c}{$t$-stat} \\
  \colrule
 $000001$ & $0$&$062^{\ast}$ & $6.91$  & $-0$&$026^{\ast}$ & $-3.67$ & $-0$&$008$        & $-1.17$ & $0$&$000$         & $0.05$  & $0$&$000$         & $0.12$  & $0$&$002$  & $0.26$ \\%
 $000002$ & $0$&$094^{\ast}$ & $24.02$ & $-0$&$034^{\ast}$ & $-11.58$& $-0$&$014^{\ast}$ & $-4.38$ & $-0$&$009^{\ast}$ & $-3.09$ & $0$&$004$         & $1.34$  & $0$&$005$  & $1.77$\\%
 $000009$ & $0$&$242$        & $1.23$  & $-0$&$545$        & $-1.74$ & $0$&$277$         & $1.86$  & $-0$&$334$        & $-1.17$ & $-0$&$053$        & $-0.23$ & $-0$&$157$ & $-0.83$\\%
 $000012$ & $0$&$077$        & $1.43$  & $0$&$023$         & $0.80$  &    &              &          & $-0$&$020$        & $-1.28$ & $0$&$035$        & $1.05$  & $-0$&$017$ & $-0.75$\\%
 $000016$ & $0$&$134^{\ast}$ & $13.01$ & $-0$&$013^{\ast}$ & $-2.59$ & $-0$&$007$        & $-1.48$ & $-0$&$019^{\ast}$ & $-2.72$ & $-0$&$007$        & $-0.19$ & $-0$&$009$ & $-1.43$\\%
 $000021$ & $0$&$132^{\ast}$ & $8.39$  & $-0$&$005$        & $-0.46$ & $-0$&$009$        & $-0.88$ & $-0$&$012$        & $-1.06$ & $0$&$002$         & $0.19$  & $-0$&$008$ & $-0.74$\\%
 $000024$ & $0$&$131^{\ast}$ & $14.90$ & $-0$&$020^{\ast}$ & $-3.62$ & $-0$&$011^{\ast}$ & $-2.20$ & $-0$&$002$        & $-0.70$ &$-0$&$009^{\ast}$ &$-2.09$ & $0$&$003$  & $0.56$\\%
 $000027$ & $0$&$073^{\ast}$ & $5.48$  & $-0$&$022^{\ast}$ & $-2.32$ & $0$&$012$         & $1.29$  & $-0$&$007$        & $-0.77$ & $0$&$001$         & $0.16$  & $0$&$022^{\ast}$  & $2.10$\\%
 $000063$ & $0$&$119^{\ast}$ & $15.51$ & $-0$&$009$        & $-1.88$ & $-0$&$008^{\ast}$ & $-2.11$ & $-0$&$003$        & $-0.63$ & $-0$&$002$        & $-0.64$ & $-0$&$001$ & $-0.21$\\%
 $000066$ & $0$&$136^{\ast}$ & $6.09$  & $-0$&$018^{\ast}$ & $-2.98$ & $0$&$003$         & $0.57$  & $-0$&$011$        & $-0.96$ & $-0$&$006$        & $-0.71$ & $0$&$003$  & $0.40$\\%
 $000088$ & $0$&$125^{\ast}$ & $26.45$ & $-0$&$003$        & $-1.50$ & $-0$&$007^{\ast}$ & $-4.16$ & $-0$&$008^{\ast}$ & $-4.39$ & $-0$&$004$        & $-1.58$ & $0$&$002$  & $1.45$\\%
 $000089$ & $0$&$100^{\ast}$ & $22.83$ & $-0$&$030^{\ast}$ & $-8.37$ & $-0$&$009^{\ast}$ & $-3.30$ & $-0$&$002$        & $-0.76$ & $0$&$003$         & $1.27$  & $0$&$002$  & $1.00$\\%
 $000406$ & $0$&$105^{\ast}$ & $6.15$  & $-0$&$015$        & $-1.87$ & $-0$&$022$        & $-1.73$ & $-0$&$008$        & $-0.89$ & $0$&$020$         & $1.60$  & $-0$&$006$ & $-0.73$\\%
 $000429$ & $0$&$105$    & $0.64$      & $-0$&$282$        & $-0.86$ & $-0$&$105$        & $-0.64$ &     &             &         &    &              &          & $0$&$166$  & $0.65$\\%
 $000488$ & $0$&$128^{\ast}$ & $31.92$ & $-0$&$026^{\ast}$ & $-11.70$& $-0$&$015^{\ast}$ & $-6.66$ & $-0$&$003$        & $-1.60$ & $-0$&$005$        & $-1.95$ & $-0$&$003$ & $-1.60$\\%
 $000539$ & $0$&$107^{\ast}$ & $14.42$ & $-0$&$031^{\ast}$ & $-6.30$ & $-0$&$004$        & $-0.91$ & $-0$&$001$        & $-0.17$ & $-0$&$005$        & $-1.19$ & $-0$&$003$ & $-0.64$\\%
 $00054$ & $0$&$139^{\ast}$ & $3.63$  & $-0$&$013$        & $-0.78$ & $-0$&$014$        & $-1.22$ & $-0$&$016$        & $-0.54$ & $0$&$011$         & $1.37$  & $0$&$023$  & $1.10$\\%
 $000550$ & $1$&$403$    & $2.39$      &     &             &         &     &             &          &     &             &          &    &            &          &  &         &  \\%
 $000581$ & $0$&$115^{\ast}$ & $21.53$ & $-0$&$010^{\ast}$ & $-3.63$ & $-0$&$008^{\ast}$ & $-3.19$ & $-0$&$005$        & $-1.90$ &$-0$&$005^{\ast}$  & $-2.28$ & $-0$&$005^{\ast}$ & $-1.98$\\%
 $000625$ & $0$&$085^{\ast}$ & $18.60$ & $-0$&$025^{\ast}$ & $-6.91$ & $-0$&$021^{\ast}$ & $-5.15$ & $-0$&$004$        & $-1.13$ & $-0$&$004$        & $-1.04$ & $0$&$000$ & $-0.07$\\%
 $000709$ & $0$&$098^{\ast}$ & $11.60$ & $-0$&$031^{\ast}$ & $-5.35$ & $-0$&$013$        & $-1.89$ & $-0$&$006$        & $-1.18$ &$-0$&$014^{\ast}$  & $-2.46$ & $0$&$002$  & $0.28$\\%
 $000720$ & $0$&$162^{\ast}$ & $13.47$ & $-0$&$015^{\ast}$ & $-2.04$ & $0$&$009^{\ast}$  & $1.99$  & $-0$&$008$        & $-1.02$ & $0$&$003$         & $0.48$  & $0$&$003$  & $0.46$\\%
 $000778$ & $0$&$130^{\ast}$ & $5.06$  & $-0$&$015$        & $-1.52$ & $-0$&$010$        & $-1.43$ & $-0$&$004$        & $-0.62$ & $-0$&$013$        & $-1.07$ & $-0$&$005$ & $-0.63$\\%
  \botrule
  \end{tabular}}
\label{TB:impact:tem1:ins}
\end{minipage}
\end{center}
\end{table}

The temporary impact of transactions inside trade packages for individuals persists over a horizon longer than the impact time for institutions. In Table~\ref{TB:impact:tem1:ind}, the estimated coefficients $\beta_i$ with $i=0,5,10,15,20,25$ seconds using the records of transactions inside trade packages executed as market orders from individuals are depicted. Coefficient $\beta_{0sec}$ is positive and significant for all the 23 stocks, which further verifies the strong impact of the trading volume on the current price return. Moreover, $\beta_{0sec}$ is larger than the magnitudes of coefficients $\beta_i$ with $i>0$. Coefficients $\beta_i$ with time lag $5 \leq i \leq 20$ seconds are negative and significant for most stocks, and $\beta_{25sec}$ is significant for almost half of the stocks. This implies that the trading volumes of transactions inside trade packages for individuals have temporary negative effects on price returns for about $20-25$ seconds. Similar to the trading volumes executed as limited orders from institutions, the trading volumes executed as limited orders from individuals have a negative impact on current price returns, and the following price returns tend to reverse soon.

\begin{table}
\begin{center}
\begin{minipage}[c]{\linewidth}
  \tbl{Coefficients and $t$-statistics for the linear fit of Equation~(\ref{Eq:impact:tem1}) using the records of transactions inside trade packages executed as market orders from individuals. Coefficients marked with $^{\ast}$ are statistically significant at 5\% level.}
{\begin{tabular}{C{0.9cm}r@{.}lR{0.9cm}r@{.}lR{0.9cm}r@{.}lR{0.9cm}r@{.}lR{0.9cm}r@{.}lR{0.9cm}r@{.}lR{0.9cm}}
  \toprule
   Code &  \multicolumn{2}{c}{$\beta_{0}$} &  \multicolumn{1}{c}{$t$-stat} &  \multicolumn{2}{c}{$\beta_{5}$} &  \multicolumn{1}{c}{$t$-stat} &  \multicolumn{2}{c}{$\beta_{10}$} &  \multicolumn{1}{c}{$t$-stat} &  \multicolumn{2}{c}{$\beta_{15}$} &  \multicolumn{1}{c}{$t$-stat} &  \multicolumn{2}{c}{$\beta_{20}$} &  \multicolumn{1}{c}{$t$-stat} &  \multicolumn{2}{c}{$\beta_{25}$} &  \multicolumn{1}{c}{$t$-stat} \\
  \colrule
 $000001$ & $0$&$090^{\ast}$ &$104.65$ & $-0$&$042^{\ast}$ & $-53.52$ & $-0$&$011^{\ast}$ & $-14.14$ & $-0$&$003^{\ast}$ & $-3.91$ & $-0$&$001$        & $-0.83$ & $-0$&$003^{\ast}$ & $-3.52$ \\%
 $000002$ & $0$&$084^{\ast}$ & $91.13$ & $-0$&$030^{\ast}$ & $-40.15$ & $-0$&$009^{\ast}$ & $-11.54$ & $-0$&$004^{\ast}$ & $-4.77$ & $-0$&$001$        & $-1.81$ & $0$&$000$         & $0.43$\\%
 $000009$ & $0$&$096^{\ast}$ & $67.99$ & $-0$&$032^{\ast}$ & $-29.81$ & $-0$&$010^{\ast}$ & $-8.61$  & $-0$&$007^{\ast}$ & $-5.40$ & $-0$&$003^{\ast}$ & $-2.83$ & $-0$&$001$        & $-0.87$\\%
 $000012$ & $0$&$085^{\ast}$ & $91.55$ & $-0$&$026^{\ast}$ & $-36.16$ & $-0$&$013^{\ast}$ & $-17.23$ & $-0$&$007^{\ast}$ & $-9.83$ & $-0$&$003^{\ast}$ & $-4.72$ & $0$&$011^{\ast}$  & $14.90$\\%
 $000016$ & $0$&$110^{\ast}$ & $58.77$ & $-0$&$020^{\ast}$ & $-19.54$ & $-0$&$014^{\ast}$ & $-12.00$ & $-0$&$005^{\ast}$ & $-4.70$ & $-0$&$004^{\ast}$ & $-4.02$  & $0$&$000$         & $-0.26$\\%
 $000021$ & $0$&$110^{\ast}$ & $74.10$ & $-0$&$029^{\ast}$ & $-27.54$ & $-0$&$011^{\ast}$ & $-10.63$ & $-0$&$006^{\ast}$ & $-5.82$ & $-0$&$003^{\ast}$ & $-3.05$ & $-0$&$002$        & $-1.47$\\%
 $000024$ & $0$&$112^{\ast}$ & $50.20$ & $-0$&$015^{\ast}$ & $-12.02$ & $-0$&$011^{\ast}$ & $-8.12$  & $-0$&$009^{\ast}$ & $-6.83$ & $-0$&$003^{\ast}$ & $-1.97$ & $-0$&$004^{\ast}$ & $-3.09$\\%
 $000027$ & $0$&$099^{\ast}$ & $90.69$ & $-0$&$026^{\ast}$ & $-31.90$ & $-0$&$012^{\ast}$ & $-15.75$ & $-0$&$005^{\ast}$ & $-6.32$ & $-0$&$003^{\ast}$ & $-3.94$ & $0$&$002^{\ast}$  & $2.83$\\%
 $000063$ & $0$&$058^{\ast}$ & $62.64$ & $-0$&$014^{\ast}$ & $-17.99$ & $-0$&$004^{\ast}$ & $-9.63$  & $-0$&$003^{\ast}$ & $-3.80$ & $-0$&$001$        & $0.71$  & $0$&$000$         & $0.39$\\%
 $000066$ & $0$&$111^{\ast}$ & $53.63$ & $-0$&$028^{\ast}$ & $-21.86$ & $-0$&$013^{\ast}$ & $-5.54$  & $-0$&$008^{\ast}$ & $-6.22$ & $-0$&$006^{\ast}$ & $-4.33$ & $-0$&$004^{\ast}$ & $-2.95$\\%
 $000088$ & $0$&$122^{\ast}$ & $51.29$ & $-0$&$008^{\ast}$ & $-8.51$  & $-0$&$005^{\ast}$ & $-10.35$ & $-0$&$004^{\ast}$ & $-4.09$ & $-0$&$004^{\ast}$ & $-3.92$ & $0$&$001$         & $0.71$\\%
 $000089$ & $0$&$107^{\ast}$ & $75.79$ & $-0$&$016^{\ast}$ & $-18.33$ & $-0$&$009^{\ast}$ & $-13.92$ & $-0$&$005^{\ast}$ & $-5.75$ & $-0$&$003^{\ast}$ & $-3.26$ & $0$&$002^{\ast}$  & $2.03$\\%
 $000406$ & $0$&$111^{\ast}$ & $75.14$ & $-0$&$030^{\ast}$ & $-28.31$ & $-0$&$015^{\ast}$ & $-7.39$  & $-0$&$009^{\ast}$ & $-8.87$ & $-0$&$003^{\ast}$ & $-3.03$ & $0$&$000$         & $-0.18$\\%
 $000429$ & $0$&$094^{\ast}$ & $47.67$ & $-0$&$023^{\ast}$ & $-18.50$ & $-0$&$010^{\ast}$ & $-20.51$ & $-0$&$006^{\ast}$ & $-4.59$ & $-0$&$003^{\ast}$ & $-1.99$ & $-0$&$003^{\ast}$ & $-2.64$\\%
 $000488$ & $0$&$126^{\ast}$ &$107.67$ & $-0$&$022^{\ast}$ & $-36.86$ & $-0$&$013^{\ast}$ & $-12.49$ & $-0$&$006^{\ast}$ & $-10.51$& $-0$&$003^{\ast}$ & $-5.20$ & $-0$&$002^{\ast}$ & $-2.66$\\%
 $000539$ & $0$&$102^{\ast}$ & $75.63$ & $-0$&$022^{\ast}$ & $-25.49$ & $-0$&$011^{\ast}$ & $-5.05$  & $-0$&$004^{\ast}$ & $-4.53$ & $-0$&$004^{\ast}$ & $-4.73$ & $-0$&$003^{\ast}$ & $-3.47$\\%
 $000541$ & $0$&$130^{\ast}$ & $50.71$ & $-0$&$011^{\ast}$ & $-9.72$  & $-0$&$006^{\ast}$ & $-12.55$ & $-0$&$005^{\ast}$ & $-4.70$ & $-0$&$003^{\ast}$ & $-2.61$ & $-0$&$002$       & $-1.91$\\%
 $000550$ & $1$&$107^{\ast}$ & $87.79$ & $-0$&$026^{\ast}$ & $-28.81$ & $-0$&$011^{\ast}$ & $-8.81$  & $-0$&$008^{\ast}$ & $-9.20$ & $0$&$001$         & $1.57$  & $-0$&$001$       & $-0.79$\\%
 $000581$ & $0$&$126^{\ast}$ & $62.91$ & $-0$&$012^{\ast}$ & $-12.50$ & $-0$&$009^{\ast}$ & $-3.19$  & $-0$&$008^{\ast}$ & $-7.76$ & $-0$&$003^{\ast}$ & $-3.49$ & $-0$&$003^{\ast}$& $-2.82$\\%
 $000625$ & $0$&$092^{\ast}$ & $92.89$ & $-0$&$028^{\ast}$ & $-36.80$ & $-0$&$012^{\ast}$ & $-16.37$ & $-0$&$006^{\ast}$ & $-7.97$ & $-0$&$001$        & $-1.94$ & $-0$&$002^{\ast}$& $-3.02$\\%
 $000709$ & $0$&$096^{\ast}$ & $64.89$ & $-0$&$020^{\ast}$ & $-20.59$ & $-0$&$015^{\ast}$ & $-13.10$ & $-0$&$008^{\ast}$ & $-8.90$ & $-0$&$004^{\ast}$ & $-3.85$ & $0$&$001$        & $0.87$\\%
 $000720$ & $0$&$150^{\ast}$ & $89.91$ & $-0$&$012^{\ast}$ & $-12.02$ & $-0$&$005^{\ast}$ & $-5.07$  & $-0$&$002^{\ast}$ & $-2.13$ & $0$&$001$         & $0.95$  & $0$&$002$        & $1.55$\\%
 $000778$ & $0$&$105^{\ast}$ & $63.38$ & $-0$&$020^{\ast}$ & $-17.23$ & $-0$&$006^{\ast}$ & $-6.30$  & $-0$&$005^{\ast}$ & $-5.47$ & $-0$&$003^{\ast}$ & $-3.75$ & $0$&$000$        & $0.31$\\%
  \botrule
  \end{tabular}}
\label{TB:impact:tem1:ind}
\end{minipage}
\end{center}
\end{table}

\subsubsection{Modeling price returns}

It has been shown that the trading volumes of transactions inside trade packages have a significant impact on price returns. We model the price return $R(t)$ taking into account both the autoregressive components of previous returns $R(t-j)$ and the previous trading volumes $v(t-i)$ within trade packages. The regression from is
\begin{equation}
   R(t) = \beta_0^{\ast} + \sum_{j=5}^J \gamma_j R(t-j) + \sum_{i=0}^I \beta_i s \ln v(t-i)+\epsilon(t),
   \label{Eq:impact:tem2}
\end{equation}
where $\ln v(t-i)=0$ when there is no transaction inside trade package executed at time $t-i$. We do not separately consider the effects of trading volumes of the transactions concurrently traded with those transactions inside trade packages, but contribute their impact to the corresponding price returns. The coefficient $\beta_i$ reflects the response of the return $R(t)$ to the trading volume of transaction inside trade package executed $i$ seconds before time $t$. We have verified that the trading volumes of transactions inside trade packages have a temporary impact on price returns for less than half a minute, and therefore we restrict $I=25$ seconds. The parameter $J$ is also set to $J=25$ seconds to simplify the regression.

We report the estimated coefficients $\beta_i$ by fitting Equation~(\ref{Eq:impact:tem2}) with $i=0,5,10,15,20,25$ seconds using the records of transactions inside trade packages executed as market orders from institutions in Table~\ref{TB:impact:tem2:ins}. According to the $R^2$ in the last column, the fitted model could explain considerable portion, generally has $R^2\geq20\%$, of the variance in price returns. For $i=0$, positive and significant coefficient $\beta_{0sec}$ is observed in all the 23 stocks, displaying values significantly larger than the magnitudes of $\beta_i$ with $i>0$. This further confirms that the trading volumes of transactions inside trade packages executed as market orders have a strong impact on current price returns. However, the negative relation between the price return and the trading volume executed $5$ seconds before is weakened, not as significant as that observed in the simple regression of Equation~(\ref{Eq:impact:tem1}). After taking into account the autoregressive components, only five stocks have significant coefficients $\beta_{5sec}$, and four of them are negative. To further increase the time lag $i$, the number of stocks which have significant coefficients $\beta_i$ is even smaller. The shortening of the impact time may because the price return can be partially explained by the autoregressive components of previous returns.

\begin{table}
\begin{center}
\begin{minipage}[c]{\linewidth}
  \tbl{Coefficients and $t$-statistics for the linear fit of Equation~(\ref{Eq:impact:tem2}) with $I=25$ and $J=25$ seconds using the records of transactions inside trade packages executed as market orders from institutions. Coefficients marked with $^{\ast}$ are statistically significant at 5\% level.}
{\resizebox{1.0\linewidth}{!}{
\begin{tabular}{C{0.9cm}r@{.}lR{0.9cm}r@{.}lR{0.9cm}r@{.}lR{0.9cm}r@{.}lR{0.9cm}r@{.}lR{0.9cm}r@{.}lR{0.9cm}C{0.7cm}}
  \toprule
   Code &  \multicolumn{2}{c}{$\beta_{0}$} &  \multicolumn{1}{c}{$t$-stat} &  \multicolumn{2}{c}{$\beta_{5}$} &  \multicolumn{1}{c}{$t$-stat} &  \multicolumn{2}{c}{$\beta_{10}$} &  \multicolumn{1}{c}{$t$-stat} &  \multicolumn{2}{c}{$\beta_{15}$} &  \multicolumn{1}{c}{$t$-stat} &  \multicolumn{2}{c}{$\beta_{20}$} &  \multicolumn{1}{c}{$t$-stat} &  \multicolumn{2}{c}{$\beta_{25}$} &  \multicolumn{1}{c}{$t$-stat} & $R^2$\\
  \colrule
 $000001$ & $0$&$075^{\ast}$ & $11.63$ & $0$&$007$         & $0.96$  & $-0$&$020^{\ast}$ & $-2.87$ & $-0$&$013$       & $-1.89$ & $-0$&$012$        & $-1.66$ & $0$&$007$  & $1.04$   & $0.23$ \\%
 $000002$ & $0$&$085^{\ast}$ & $31.28$ & $-0$&$007^{\ast}$ & $-2.43$ & $-0$&$008^{\ast}$ & $-2.54$ & $-0$&$006^{\ast}$& $-2.05$ & $-0$&$001$        & $-0.48$ & $-0$&$005$ & $-1.62$ & $0.24$ \\%
 $000009$ & $0$&$027$        & $0.74$  & $-0$&$041$        & $-1.18$ & $-0$&$045$        & $-1.17$ & $-0$&$048$       & $-1.21$ & $-0$&$016$        & $-0.41$ & $-0$&$027$ & $-0.62$  & $0.17$ \\%
 $000012$ & $0$&$103^{\ast}$ & $3.76$  & $0$&$023$         & $0.72$  & $0$&$016$         & $0.52$  & $-0$&$010$       & $-0.32$ & $0$&$007$         & $0.23$  & $-0$&$002$ & $-0.06$  & $0.27$ \\%
 $000016$ & $0$&$135^{\ast}$ & $20.11$ & $0$&$002$         & $-0.25$ & $-0$&$008$        & $-1.01$ & $0$&$001$        & $0.15$  & $-0$&$012$        & $-1.40$ & $-0$&$008$ & $-0.94$  & $0.29$ \\%
 $000021$ & $0$&$137^{\ast}$ & $11.80$ & $0$&$023$         & $1.64$  & $-0$&$013$        & $-0.88$ & $-0$&$011$       & $-0.81$ & $-0$&$003$        & $-0.25$ & $-0$&$010$ & $-0.71$  & $0.29$ \\%
 $000024$ & $0$&$129^{\ast}$ & $24.81$ & $0$&$000$         & $0.06$  & $-0$&$010$        & $-1.57$ & $-0$&$005$       & $-0.80$ & $-0$&$002$        & $-0.33$ & $-0$&$005$ & $-0.69$ & $0.30$ \\%
 $000027$ & $0$&$082^{\ast}$ & $9.28$  & $0$&$011$         & $1.10$  & $-0$&$001$        & $-0.07$ & $0$&$001$        & $0.12$  & $0$&$007$         & $0.68$  & $0$&$013$  & $1.27$   & $0.20$ \\%
 $000063$ & $0$&$106^{\ast}$ & $25.72$ & $0$&$005$         & $0.92$  & $0$&$003$         & $0.56$  & $0$&$005$        & $0.96$  & $0$&$003$         & $0.60$  & $-0$&$004$ & $-0.75$  & $0.27$ \\%
 $000066$ & $0$&$154^{\ast}$ & $15.03$ & $0$&$002$         & $0.16$  & $-0$&$015$        & $-1.18$ & $-0$&$007$       & $-0.56$ & $-0$&$005$        & $-0.35$ & $-0$&$021$ & $-1.59$ & $0.31$ \\%
 $000088$ & $0$&$064^{\ast}$ & $37.48$ & $-0$&$007^{\ast}$ & $-3.62$ & $-0$&$001$        & $-1.05$ & $0$&$000$        & $0.09$  & $0$&$000$         & $-0.17$ & $0$&$003$  & $1.25$   & $0.13$ \\%
 $000089$ & $0$&$093^{\ast}$ & $35.43$ & $0$&$005$         & $1.83$  & $0$&$000$         & $-1.54$ & $0$&$002$        & $0.60$  & $0$&$002$         & $0.80$  & $-0$&$002$ & $-0.77$ & $0.29$ \\%
 $000406$ & $0$&$107^{\ast}$ & $9.87$  & $-0$&$001$        & $-0.10$ & $-0$&$014$        & $1.16$  & $0$&$006$        & $0.49$  & $0$&$000$         & $-0.04$ & $0$&$000$  & $0.02$    & $0.17$ \\%
 $000429$ & $0$&$002$        & $0.02$  & $-0$&$103$        & $-1.28$ & $-0$&$084$        & $-0.64$ & $-0$&$035$       & $-0.43$ & $-0$&$040$        & $-0.50$ & $-0$&$048$ & $-0.60$ & $0.18$ \\%
 $000488$ & $0$&$128^{\ast}$ & $54.40$ & $-0$&$004$        & $-1.66$ & $-0$&$004$        & $-0.82$ & $-0$&$002$       & $-0.81$ & $-0$&$003$        & $-1.22$ & $-0$&$004^{\ast}$& $-3.15$ & $0.23$ \\%
 $000539$ & $0$&$093^{\ast}$ & $23.32$ & $-0$&$013^{\ast}$ & $-2.83$ & $0$&$005$         & $-0.02$ & $-0$&$005$       & $-1.16$ & $-0$&$001$        & $-0.30$ & $0$&$000$  & $-0.02$   & $0.23$ \\%
 $000541$ & $0$&$161^{\ast}$ & $8.18$  & $-0$&$015$        & $-0.59$ & $-0$&$016$        & $-3.28$ & $-0$&$005$       & $-0.20$ & $0$&$022$         & $0.88$  & $-0$&$003$ & $-0.10$ & $0.34$ \\%
 $000550$ & $0$&$137$        & $1.92$  & $-0$&$085$        & $-0.76$ & $-0$&$087$        & $-0.86$ & $-0$&$087$       & $-0.83$ & $-0$&$088$         & $-0.82$ & $-0$&$085$ & $-1.15$ & $0.80$ \\%
 $000581$ & $0$&$084^{\ast}$ & $37.77$ & $-0$&$007^{\ast}$ & $-2.63$ & $0$&$000$         & $3.97$  & $-0$&$013^{\ast}$& $-4.96$ & $-0$&$008^{\ast}$ & $-2.87$ & $-0$&$004$ & $-1.32$  & $0.19$ \\%
 $000625$ & $0$&$075^{\ast}$ & $22.70$ & $-0$&$004$        & $-1.24$ & $-0$&$012^{\ast}$ & $-0.90$ & $-0$&$005$       & $-1.43$ & $-0$&$005$        & $-1.38$ & $-0$&$007$ & $-1.80$  & $0.20$ \\%
 $000709$ & $0$&$084^{\ast}$ & $16.25$ & $-0$&$011$        & $-1.84$ & $-0$&$005$        & $-1.89$ & $-0$&$013^{\ast}$& $-2.12$ & $-0$&$015^{\ast}$ & $-2.37$ & $-0$&$002$ & $-0.23$ & $0.26$ \\%
 $000720$ & $0$&$122^{\ast}$ & $20.49$ & $0$&$021^{\ast}$  & $3.05$  & $0$&$027^{\ast}$  & $1.99$  & $0$&$008$        & $1.19$  & $0$&$017^{\ast}$  & $0.50$  & $0$&$011$  & $1.52$   & $0.24$ \\%
 $000778$ & $0$&$134^{\ast}$ & $11.89$ & $-0$&$003$        & $0.20$  & $-0$&$012$        & $-1.43$ & $-0$&$014$       & $-1.04$ & $-0$&$010$        & $-0.74$ & $-0$&$007$ & $-0.52$  & $0.32$ \\%
  \botrule
  \end{tabular}}}
\label{TB:impact:tem2:ins}
\end{minipage}
\end{center}
\end{table}

We also obtain the coefficients $\beta_i$ by estimating Equation~(\ref{Eq:impact:tem2}) using the records of transactions inside trade packages executed as market orders from individuals as depicted in Table~\ref{TB:impact:tem2:ind}. The $R^2$ for individuals is generally smaller than that for institutions, but still has values $\geq10\%$ for most stocks. Coefficient $\beta_{0sec}$ for individuals is positive and significant for all the stocks, but approximately one order of magnitude smaller than $\beta_{0sec}$ for institutions. This implies that the positive impact of trading volumes on current price returns for individuals is not as strong as that for institutions. More than half of the stocks have negative and significant coefficients $\beta_i$ for $i\leq 25$ seconds. This indicates that for more than half of the stocks the trading volumes of transactions inside trade packages have a temporary negative impact for at least $25$ seconds. In comparison with the trading volumes for institutions, the trading volumes for individuals have a negative price impact over a relatively longer horizon even after considering the autoregressive components of previous returns.

\begin{table}
\begin{center}
\begin{minipage}[c]{\linewidth}
  \tbl{Coefficients and $t$-statistics for the linear fit of Equation~(\ref{Eq:impact:tem2}) with $I=25$ and $J=25$ seconds using the records of transactions inside trade packages executed as market orders from individuals. Coefficients marked with $^{\ast}$ are statistically significant at 5\% level.}
{\resizebox{1.0\linewidth}{!}{
\begin{tabular}{C{0.9cm}r@{.}lR{0.9cm}r@{.}lR{0.9cm}r@{.}lR{0.9cm}r@{.}lR{0.9cm}r@{.}lR{0.9cm}r@{.}lR{0.9cm}C{0.7cm}}
  \toprule
   Code &  \multicolumn{2}{c}{$\beta_{0}$} &  \multicolumn{1}{c}{$t$-stat} &  \multicolumn{2}{c}{$\beta_{5}$} &  \multicolumn{1}{c}{$t$-stat} &  \multicolumn{2}{c}{$\beta_{10}$} &  \multicolumn{1}{c}{$t$-stat} &  \multicolumn{2}{c}{$\beta_{15}$} &  \multicolumn{1}{c}{$t$-stat} &  \multicolumn{2}{c}{$\beta_{20}$} &  \multicolumn{1}{c}{$t$-stat} &  \multicolumn{2}{c}{$\beta_{25}$} &  \multicolumn{1}{c}{$t$-stat} & $R^2$\\
  \colrule
 $000001$ & $0$&$008^{\ast}$ & $65.42$  & $-0$&$002^{\ast}$ & $-11.77$ & $-0$&$001^{\ast}$ & $-8.69$ & $-0$&$001^{\ast}$ & $-6.01$ & $-0$&$002^{\ast}$ & $-14.38$& $-0$&$001^{\ast}$& $-5.37$  & $0.19$ \\%
 $000002$ & $0$&$012^{\ast}$ & $58.75$  & $-0$&$006^{\ast}$ & $-25.44$ & $0$&$000$         & $1.64$  & $0$&$001^{\ast}$  & $2.04$  & $-0$&$001^{\ast}$ & $-3.09$ & $0$&$000$        & $-0.67$  & $0.15$ \\%
 $000009$ & $0$&$015^{\ast}$ & $55.03$  & $-0$&$004^{\ast}$ & $-10.95$ & $0$&$001^{\ast}$  & $2.64$  & $0$&$000$         & $1.14$  & $-0$&$002^{\ast}$ & $-7.02$ & $-0$&$002^{\ast}$& $-5.87$  & $0.14$ \\%
 $000012$ & $0$&$044^{\ast}$ & $108.68$ & $-0$&$008^{\ast}$ & $-19.14$ & $-0$&$003^{\ast}$ & $-6.01$ & $-0$&$003^{\ast}$ & $-7.61$ & $-0$&$002^{\ast}$ & $-4.30$ & $-0$&$003^{\ast}$& $-6.01$  & $0.20$ \\%
 $000016$ & $0$&$079^{\ast}$ & $85.62$  & $-0$&$004^{\ast}$ & $-3.56$  & $-0$&$005^{\ast}$ & $-4.69$ & $-0$&$002^{\ast}$ & $-2.31$ & $-0$&$005^{\ast}$ & $-5.54$ & $-0$&$006^{\ast}$& $-6.03$  & $0.19$ \\%
 $000021$ & $0$&$012^{\ast}$ & $48.69$  & $-0$&$003^{\ast}$ & $-10.68$ & $0$&$000$         & $-1.03$ & $0$&$000$         & $1.48$  & $-0$&$002^{\ast}$ & $-6.97$ & $-0$&$004^{\ast}$& $-14.15$ & $0.11$ \\%
 $000024$ & $0$&$021^{\ast}$ & $35.31$  & $-0$&$008^{\ast}$ & $-11.68$ & $-0$&$003^{\ast}$ & $-4.00$ & $0$&$002^{\ast}$  & $3.30$  & $0$&$002^{\ast}$  & $2.42$  & $-0$&$002^{\ast}$& $-3.09$  & $0.09$ \\%
 $000027$ & $0$&$007^{\ast}$ & $50.89$  & $-0$&$004^{\ast}$ & $-26.08$ & $0$&$002^{\ast}$  & $13.20$ & $-0$&$002^{\ast}$ & $-14.16$& $0$&$000^{\ast}$  & $-2.31$ & $0$&$000^{\ast}$ & $-2.64$  & $0.11$ \\%
 $000063$ & $0$&$009^{\ast}$ & $62.02$  & $-0$&$003^{\ast}$ & $-19.95$ & $-0$&$001^{\ast}$ & $-7.09$ & $0$&$000^{\ast}$  & $-2.08$ & $0$&$000^{\ast}$  & $1.98$  & $-0$&$001^{\ast}$& $-6.86$  & $0.17$ \\%
 $000066$ & $0$&$042^{\ast}$ & $55.68$  & $-0$&$008^{\ast}$ & $-9.32$  & $-0$&$004^{\ast}$ & $-4.33$ & $-0$&$003^{\ast}$ & $-3.28$ & $-0$&$004^{\ast}$ & $-5.09$ & $-0$&$001$       & $-1.60$  & $0.14$ \\%
 $000088$ & $0$&$003^{\ast}$ & $26.57$  & $-0$&$002^{\ast}$ & $-11.87$ & $0$&$000$         & $0.26$  & $0$&$000$         & $0.22$  & $0$&$000$         & $-1.86$ & $-0$&$001^{\ast}$& $-6.70$  & $0.03$ \\%
 $000089$ & $0$&$012^{\ast}$ & $54.11$  & $-0$&$001^{\ast}$ & $-2.27$  & $0$&$001^{\ast}$  & $2.69$  & $0$&$000$         & $-0.05$ & $0$&$001^{\ast}$  & $4.00$  & $0$&$001^{\ast}$ & $2.14$   & $0.09$ \\%
 $000406$ & $0$&$027^{\ast}$ & $61.79$  & $-0$&$006^{\ast}$ & $-11.87$ & $-0$&$002^{\ast}$ & $-3.429$& $-0$&$002^{\ast}$ & $-5.13$ & $0$&$000$         & $0.33$  & $-0$&$001$       & $-1.14$  & $0.15$ \\%
 $000429$ & $0$&$074^{\ast}$ & $73.05$  & $-0$&$007^{\ast}$ & $-5.42$  & $-0$&$003^{\ast}$ & $-2.59$ & $-0$&$005^{\ast}$ & $-4.33$ & $-0$&$007^{\ast}$ & $-5.83$ & $-0$&$007^{\ast}$& $-5.51$  & $0.18$ \\%
 $000488$ & $0$&$127^{\ast}$ & $208.16$ & $-0$&$006^{\ast}$ & $-8.84$  & $-0$&$005^{\ast}$ & $-7.06$ & $0$&$000$         & $-0.35$ & $-0$&$003^{\ast}$ & $-3.93$ & $-0$&$005^{\ast}$& $-7.69$  & $0.25$ \\%
 $000539$ & $0$&$005^{\ast}$ & $106.43$ & $-0$&$002^{\ast}$ & $-30.85$ & $0$&$001^{\ast}$  & $8.60$  & $-0$&$001^{\ast}$ & $-10.32$& $0$&$000^{\ast}$  & $-5.03$ & $0$&$000^{\ast}$ & $-6.55$  & $0.14$ \\%
 $000541$ & $0$&$105^{\ast}$ & $81.61$  & $-0$&$006^{\ast}$ & $-4.31$  & $-0$&$009^{\ast}$ & $-5.60$ & $-0$&$013^{\ast}$ & $-8.14$ & $-0$&$008^{\ast}$ & $-5.10$ & $-0$&$007^{\ast}$& $-4.50$  & $0.19$ \\%
 $000550$ & $0$&$005^{\ast}$ & $51.66$  & $-0$&$002^{\ast}$ & $-15.97$ & $-0$&$001^{\ast}$ & $-4.99$ & $0$&$000$         & $0.71$  & $0$&$000$         & $1.33$  & $-0$&$001^{\ast}$& $-4.10$  & $0.09$ \\%
 $000581$ & $0$&$005^{\ast}$ & $32.87$  & $-0$&$003^{\ast}$ & $-17.42$ & $0$&$001^{\ast}$  & $2.64$  & $-0$&$001^{\ast}$ & $-3.68$ & $0$&$000$         & $1.29$  & $-0$&$001^{\ast}$& $-6.79$  & $0.05$ \\%
 $000625$ & $0$&$006^{\ast}$ & $60.96$  & $-0$&$001^{\ast}$ & $-9.98$  & $0$&$000$         & $1.50$  & $0$&$001$         & $0.50$  & $0$&$000$         & $-0.27$ & $0$&$000^{\ast}$ & $-3.25$  & $0.14$ \\%
 $000709$ & $0$&$008^{\ast}$ & $48.83$  & $-0$&$002^{\ast}$ & $-8.36$  & $0$&$000$         &$-1.49$  & $-0$&$002^{\ast}$ & $-9.89$ & $0$&$001^{\ast}$  & $2.57$  & $0$&$000$        & $-0.90$  & $0.15$ \\%
 $000720$ & $0$&$019^{\ast}$ & $76.99$  & $-0$&$002^{\ast}$ & $-6.94$  & $0$&$001$         & $1.98$  & $-0$&$001^{\ast}$ & $-4.89$ & $0$&$001^{\ast}$  & $5.08$  & $0$&$000$        & $-1.54$  & $0.08$ \\%
 $000778$ & $0$&$055^{\ast}$ & $77.13$  & $0$&$000$         & $-0.29$  & $0$&$001$         & $0.86$  & $-0$&$002$        & $-1.92$ & $-0$&$003^{\ast}$ & $-4.54$ & $0$&$001$        & $0.84$   & $0.17$ \\%
  \botrule
  \end{tabular}}}
\label{TB:impact:tem2:ind}
\end{minipage}
\end{center}
\end{table}

\section{Summary}
\label{sec:sum}

This paper studies the characteristic properties of trade packages made by both institutional and individual investors trading on Shenzhen Stock Exchange. The sequence of transactions are grouped into packages with mostly buy or sell trades separated by the break of $1,5,10$ days. The probability distributions of the variables, i.e., the execution time $T$, the number of trades $N$ and the total trading volume $V$, characterizing trade packages show power-law tails, and a power-law fitting method based on $KS$ statistic is adopted to estimate their exponents. The exponent of the execution time is smaller than 1.0, and the exponent of the trades number is around 3.0 close to that of the total trading volume. The exponents vary with the break length, and tend to be constant for the break length longer than 5 days. Moreover, the exponents for individuals are slightly larger than those for institutions. The scaling relations between these variables are also detected. All the scaling exponents show values smaller than 1.0, different from those found by \citet{Vaglica-Lillo-Moro-Mantegna-2008-PRE} and \citet{Moro-Vicente-Moyano-Gerig-Farmer-Vaglica-Lillo-2009-PRE}.

We have also studied the trading profile of the trade packages accomplished within one day. By investigating the mean trading volume of the individual transactions inside trade packages, we find that large amounts of shares are more likely to be executed close to the opening and closing time respectively for institutions and individuals. The probability distribution of the transaction time shows that both institutional and individual investors trade more frequently after the midday break, and it displays a maximum at the closing time of the day. This phenomena might be explained by their preferences on the initial and final time of trade packages. The profile of the total trading volume further implies that institutions may be more informed than individuals.

The price impact of trade packages and its relations with the execution time $T$ and the total trading volume $V$ are analyzed respectively. We find the price impact remains nonzero over the whole period of execution time. The price impact has an empirical power-law relation with the total trading volume. Furthermore, the instantaneous price impact of the transactions inside trade packages is investigated, and transactions close to the opening and closing time of the day have a stronger impact than those during the remainder of the day. Similar to trade packages, the transactions inside trade packages also show a power-law dependence on their corresponding volumes. To study the temporary impact of transactions inside trade packages, we regress the price return against its current and previous trading volumes within trade packages, and further model the price return taking into account both the autoregressive components of previous returns and the current and previous trading volumes within trade packages. The regression results show that the trading volumes of the transactions inside trade packages have a strong positive impact on current price returns, while the reversals of the following prices are less strong and persist over only a few seconds. This may help explain the permanent impact of isolated transactions revealed in the empirical study of block trades \citep{Keim-Madhavan-1996-RFS,Gregoriou-2008-JES,Kraus-Stoll-1972-JF,Gemmill-1996-JF}. In addition, the impact of trading volumes on current price returns for institutions is stronger than that for individuals, and the price reversal for individuals persists over a horizon relatively longer than that for institutions. Though the transactions inside trade packages have a temporary impact persisting over only a few seconds, the cumulative impact of trade packages is significant over the whole period of execution time \citep{Gemmill-1996-JF}.

\section*{Acknowledgements}
We are grateful to Zhi-qiang Jiang (School of Business, East China University of Science and Technology) for preprocessing the data analyzed in this work. This work was partially supported by the National Natural Science Foundation (No. 10905023), the Zhejiang Provincial Natural Science Foundation of China (No. Z6090130), Humanities and Social Sciences Fund sponsored by Ministry of Education of the People's Republic of China (No. 09YJCZH042), and the Fundamental Research Funds for the Central Universities.

\bibliographystyle{rQUF}
\bibliography{E:/Papers/Auxiliary/Bibliography}
\vspace{36pt}

\end{document}